\documentclass[english,twocolumn,8pt]{article}
\usepackage{pdfpages}
\usepackage[utf8]{inputenc}
\usepackage[T1]{fontenc}
\usepackage{cite}
\usepackage{import, xspace, url, breakurl, capt-of, titling, abstract, mathtools}
\usepackage{amsfonts,braket,times,graphicx,color,siunitx,enumitem,authblk,lmodern,qcircuit}
\usepackage{epstopdf}
\usepackage{adjustbox}
\usepackage[labelfont=bf]{caption}
\usepackage[caption=false]{subfig}
\usepackage{textcomp}
\usepackage{siunitx}
\usepackage[section]{placeins}
\usepackage[breaklinks]{hyperref}
\usepackage[dvipsnames]{xcolor}
\usepackage[switch]{lineno}
\modulolinenumbers[5]  

\usepackage[normalem]{ulem}

\renewcommand{\figurename}{Fig.}

\DeclareCaptionLabelSeparator{bar}{ | }
\title{Operating two exchange-only qubits in parallel}
\author[1]{ Mateusz T. M\k{a}dzik\thanks{mateusz.madzik@intel.com}$^{\dagger}$} 
\author[1]{ Florian Luthi\thanks{These authors contributed equally}} 
\author[1]{Gian Giacomo Guerreschi} 
\author[1]{Fahd A. Mohiyaddin} 
\author[1]{Felix Borjans} 
\author[1]{Jason D. Chadwick} 
\author[1]{Matthew J. Curry} 
\author[1]{Joshua Ziegler} 

\author[1]{Sarah Atanasov} 
\author[1]{Peter L. Bavdaz} 
\author[1]{Elliot J. Connors} 
\author[1]{J. Corrigan} 
\author[1]{H. Ekmel Ercan} 
\author[1]{Robert Flory} 
\author[1]{Hubert C. George} 
\author[1]{Benjamin Harpt} 
\author[1]{Eric Henry} 
\author[1]{Mohammad M. Islam} 
\author[1]{Nader Khammassi} 
\author[1]{Daniel Keith} 
\author[1]{Lester F. Lampert} 
\author[1]{Todor M. Mladenov}
\author[1]{Randy W. Morris} 
\author[1]{Aditi Nethwewala} 
\author[1]{Samuel Neyens} 
\author[1]{Ren\'e Otten} 
\author[1]{Linda P. Osuna Ibarra} 
\author[1]{Bishnu Patra}
\author[1]{Ravi Pillarisetty} 
\author[1]{Shavindra Premaratne} 
\author[1]{Mick Ramsey} 
\author[1]{Andrew Risinger} 
\author[1]{John Rooney} 
\author[1]{Rostyslav Savytskyy} 
\author[1]{Thomas F. Watson} 
\author[1]{Otto K. Zietz} 
\author[1]{Anne Y. Matsuura} 
\author[1]{Stefano Pellerano} 
\author[1]{Nathaniel C. Bishop} 
\author[1]{Jeanette Roberts} 
\author[1]{James S. Clarke} 

\affil[1]{Intel Corporation, Technology Research Group, Hillsboro, OR 97124, USA}

\begin{document} 
\twocolumn[
  \maketitle             
  \textbf{
  Semiconductors are among the most promising platforms to implement large-scale quantum computers, as advanced manufacturing techniques allow fabrication of large quantum dot arrays~\cite{neyens2024probing}.
  Various qubit encodings can be used to store and manipulate quantum information on these quantum dot arrays.
  Regardless of qubit encoding, precise control over the exchange interaction between electrons confined in quantum dots in the array is critical. 
  Furthermore, it is necessary to execute high-fidelity quantum operations concurrently to make full use of the limited coherence of individual qubits.
  Here, we demonstrate the parallel operation of two exchange-only qubits, consisting of six quantum dots in a linear arrangement.
  Using randomized benchmarking techniques, we show that issuing pulses on the five barrier gates to modulate exchange interactions in a maximally parallel way maintains the quality of qubit control relative to sequential operation.
  The techniques developed to perform parallel exchange pulses can be readily adapted to other quantum-dot based encodings.
  Moreover, we show the first experimental demonstrations of an iSWAP gate and of a charge-locking Pauli spin blockade readout method.
  The results are validated using cross-entropy benchmarking~\cite{arute2019quantum}, a technique useful for performance characterization of larger quantum computing systems; here it is used for the first time on a quantum system based on semiconductor technology.}
  \\]
\saythanks

\nolinenumbers

Parallelizing operations is one of the key enablers for quantum computation with many qubits, regardless of the technological platform. 
It minimizes qubit idle time which leads to decoherence and information loss. 
While parallel operations are important for running near-term algorithms within the device's coherence time, they become crucial for effective error correction to enable fault-tolerant computation~\cite{shor1995scheme, google2023suppressing,fowler2012surface,acharya2024quantum}.
It is therefore necessary to establish experimental procedures for parallelization of qubit operations and evaluate how much parallelization is possible in order to guide workload development and future device design.

Among various quantum computing technologies, spin qubits encoded on quantum dots are emerging as a promising platform through their inherent compatibility with silicon foundries~\cite{george202412, steinacker2024300, stuyck2024cmos, zwerver2022qubits,maurand2016cmos}. 
In these devices, either electrons or holes are trapped in voltage-electrode defined quantum dot potentials and the quantum information is encoded either on a single~\cite{loss1998quantum} or multiple spins~\cite{kempe2001theory, divincenzo2000universal, koh2012pulse, levy2002universal}. 
The choice of qubit encoding determines the type of operations necessary to execute single- or two-qubit operations. 
For example, the Loss-DiVincenzo (LD) qubit encodes the quantum information onto the spin of a single electron; it relies on microwave (MW) pulses resonant with the spin's Larmor precession frequency (typically 1-\SI{20}{\giga\hertz}) to perform single qubit gates \cite{yang2019silicon, yoneda2018quantum}. 
Parallelization can then be achieved by applying two or more (e.g. frequency multiplexed) MW pulses to the sample, where special care needs to be taken to avoid crosstalk effects~\cite{lawrie2023simultaneous}.
Two-qubit gates with LD qubits are typically performed by dynamically turning on exchange coupling via voltage pulsing on barrier gates between two quantum dots \cite{reed2016reduced, xue2022quantum,mills2022two, tanttu2024assessment}. These operations so far have only been executed individually.

\begin{figure*}[!ht]
	\includegraphics[width=\textwidth]{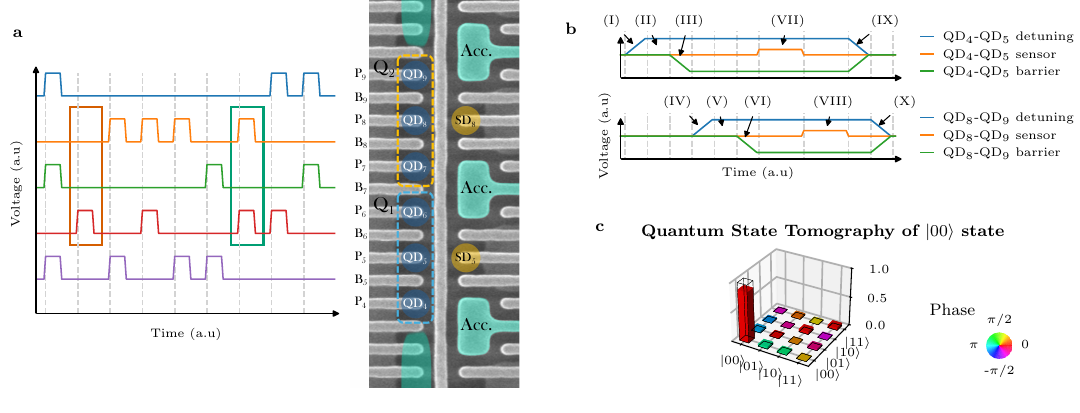}
	\caption{\textbf{Device and experimental pulse sequences}~\textbf{a)} Typical pulse sequence used in this work (left) and the Tunnel Falls device (right). The scanning electron microscope image of a device similar to the one used in this work indicates the position of quantum dots~(QD) used to form qubits (blue) and charge sensors (yellow), as well as regions of accumulated two-dimensional electron gas (green).
    The exchange-only qubits Q$_1$ and Q$_2$ are encoded on the six QDs (dashed lines).
    Voltage pulses applied sequentially (orange box) or in parallel (green box) to the electrodes dynamically control the exchange coupling strength between QDs, allowing for exchange-only qubit operations.
    \textbf{b)} Charge-locking readout sequence employed in all measurements presented in the paper (time axis not to scale). First, (I) we ramp QD$_4$-QD$_5$ to the PSB readout window in \SI{21.84}{\nano\second}, (II) subsequently allow a \SI{3.64}{\nano\second} projection time before pulsing Q$_2$ to its PSB readout window, (III, optional) before reducing the tunnel coupling between QDs involved in readout. We repeat these steps for QD$_8$-QD$_9$ in (IV), (V), and (VI). Then, (VII) the signal for QD$_4$-QD$_5$ is integrated for \SI{18}{\micro\second}. 
    Next, (VIII) we integrate the QD$_8$-QD$_9$ signal for \SI{18}{\micro\second} before ramping both qubits to their manipulation position in \SI{21.84}{\nano\second} each in (IX) and (X). 
    \textbf{c)} Quantum State Tomography (QST) of the $\ket{00}$ state. 
    We initialize by post-selection using the PSB readout sequence, and measure a fidelity of $86.4 \pm 1.4 \%$.
    }
	\label{fig:device}
\end{figure*}

Another prominent qubit encoding using quantum dots are exchange-only (EO) qubits~\cite{kempe2001theory, divincenzo2000universal}, where the quantum information is encoded on the collective spins of three electrons. 
The key appeal of this qubit type is that the single- and two-qubit gates are composed solely of voltage pulses that dynamically turn on exchange couplings $J$, eliminating the need for MW pulses.
This simplifies setup, fabrication, control and calibration requirements, especially for scalable 2-dimensional arrays. 
Recently, high fidelity single-qubit gates~\cite{andrews2019quantifying} and universal logic using two exchange-only qubits~\cite{weinstein2023universal} have both been demonstrated by applying sequential exchange pulses. Operating exchange-only qubits in parallel, however, requires simultaneous control of exchange pulses, which has only been explored without estimates of the impact on gate fidelity~\cite{fedele2021simultaneous,jirovec2025mitigation}.

In this work, we demonstrate the feasibility of simultaneous exchange control using a 12 quantum dot Tunnel Falls sample (Fig.~\ref{fig:device}\textbf{a})~\cite{george202412} (see Methods). 
We compare the system performance for sequential versus simultaneous pulsing in single- and two-qubit randomized benchmarking (RB) measurements~\cite{knill2008randomized}.
We experimentally demonstrate a new iSWAP gate implementation for exchange-only qubits, and show two-qubit gates with reduced overall duration due to parallelization of exchange pulses. 
Lastly, we implement cross-entropy benchmarking (XEB)~\cite{arute2019quantum, boixo2018characterizing} using sequential as well as simultaneous pulsing. 

\begin{figure*}[!hbt]
	\includegraphics[width=\textwidth]{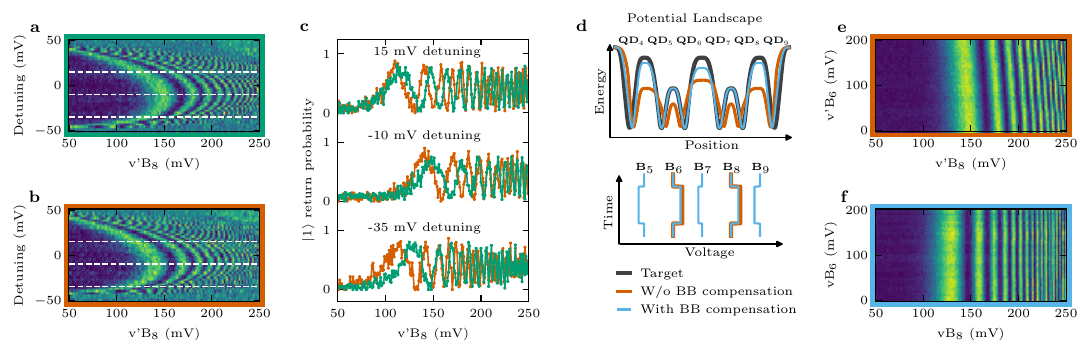}
	\caption{\textbf{Simultaneous exchange pulsing}~\textbf{a)} Exchange fingerprint for the double dot QD$_7$-QD$_8$, measured with 8 consecutive exchange pulses of \SI{10.92}{\nano\second}, each followed by a \SI{10.92}{\nano\second} pulse buffer. 
    \textbf{b)} Exchange fingerprint for the double dot QD$_7$-QD$_8$, measured in the same manner as \textbf{a)}, while also simultaneously applying a voltage $v'B_6$ on the B$_6$ electrode to emulate an exchange pulse between QD$_5$-QD$_6$. 
    \textbf{c)} Linecuts along the dashed lines of the fingerprints in \textbf{a)} and \textbf{b)}, illustrating the change in the exchange coupling due to simultaneous pulsing. 
    \textbf{d)} Qualitative diagram of potential landscape during simultaneous exchange pulses on electrodes B$_6$ and B$_8$ under different barrier-barrier compensation schemes: ideal compensation (black), no compensation (orange) and next-nearest barrier compensation (blue). Using next-nearest barrier compensation, the impact of simultaneous pulses on the potential landscape, and therefore the resulting exchange couplings, is suppressed. 
    \textbf{e)} Change in the linecut of the QD$_7$-QD$_8$ fingerprint at \SI{-10}{\milli\volt} detuning as a function of the pulse amplitude applied to B$_6$ without barrier-barrier compensation. 
    \textbf{f)} Same measurement as in \textbf{e)}, but with barrier-barrier compensation enabled. The B$_6$ amplitude dependence of the fingerprint linecut is suppressed.} 
	\label{fig:crosstalk}
\end{figure*}

We operate the central six quantum dots (QDs) of the 12 quantum dot array~\cite{george202412} to encode two exchange-only qubits Q$_1$ and Q$_2$ on dots QD$_4$-QD$_6$ and QD$_7$-QD$_9$, respectively (Fig.~\ref{fig:device}\textbf{a}, see Methods). 
To perform full two-qubit state extraction and to overcome the spin lifetime limitation of sequential readout with exchange-only qubits, we introduce charge-locking PSB readout (Fig.~\ref{fig:device}\textbf{b}, see Methods), building on the frozen PSB technique~\cite{nurizzo2023complete}. In the experiments presented here, the readout sequence is also utilized as a means of initialization through post-selection~\cite{philips2022universal}. 
We perform all exchange pulses with the same pulse duration of $t_p=\SI{10.92}{\nano\second}$, followed by a buffer time of $t_b=\SI{10.92}{\nano\second}$. 
Additionally, we use exponential overshoot pulse pre-distortions to compensate for extended wiring, limiting our signal path bandwidth, which was instrumental to achieving high-fidelity control (see Methods).
We perform state tomography on the $\ket{00}$ state to verify functionality of the system (Fig.~\ref{fig:device}\textbf{c}) and compute a fidelity of $86.4 \pm 1.4 \%$. 

\subsubsection*{Crosstalk in semiconductor qubit devices}
 
Semiconductor qubit devices experience signal crosstalk through (i), capacitive coupling of electrodes to nearby dots affecting their charge state and tunnel couplings as well as (ii), on-chip signal path routing and (iii), off-chip signal routing. 
To enable efficient control of the system, the sum of these crosstalk contributions needs to be compensated. 
The capacitive coupling crosstalk is partially corrected in a process called virtualization~\cite{hensgens2017quantum, volk2019loading}, allowing for orthogonal control over the electrochemical potentials of the QDs (see Methods). 
This reduces control complexity and ensures that the quantum dots retain their intended electron population during operation. 
In a typical spin qubit experiment, the barrier-barrier elements are left unpopulated, as all operations requiring voltage pulsing are performed sequentially. In this case, the exponential relation of the exchange coupling to the barrier electrode voltage suppresses the effects of the existing crosstalk on idle exchange axes. In this regime, a large voltage change would be needed to significancy impact the exchange coupling from its off value and thereby limit performance.
So far, barrier-barrier virtualization was realized primarily for tuning purposes~\cite{hsiao2020efficient, qiao2020coherent,jirovec2025mitigation}, but was not implemented for high-fidelity parallel exchange pulsing, which requires precise calibrations. 

We evaluate the impact of simultaneous pulses on the exchange between the QD$_7$-QD$_8$ pair (barrier electrode B$_8$)  by measuring an exchange fingerprint~\cite{reed2016reduced} (see Methods) directly~(Fig.~\ref{fig:crosstalk}\textbf{a}), and compare it to an exchange fingerprint with a simultaneous pulse of voltage $v'B_6$ on barrier electrode B$_6$~(Fig.~\ref{fig:crosstalk}\textbf{b}). 
By adding the $v'$ prefix to the voltage label, we highlight that the virtualization is incomplete and does not yet contain barrier-barrier correction factors.
We note four changes to the fingerprint: (i), a shift of the barrier voltage axis; (ii), a shift of the detuning axis; (iii) minor changes to the fingerprint shape; and (iv), a change in exchange tunability. 
The difference in the two fingerprints is visualized in Fig.~\ref{fig:crosstalk}\textbf{c}, where line-cuts of the fingerprints for 3 different detunings are shown. 
The individual exchange pulses show less exchange for the same barrier pulse voltage as compared to simultaneous pulses.
This can be understood using the diagram of the potential landscape~(Fig.~\ref{fig:crosstalk}\textbf{d}). 
The uncompensated, simultaneous exchange pulses on electrodes B$_6$ and B$_8$ lower each others potential barriers further~(Fig.~\ref{fig:crosstalk}\textbf{d}, orange curve) than the desired values~(Fig.~\ref{fig:crosstalk}\textbf{d}, black curve), therefore additionally increasing the resulting exchange coupling. Additionally, when pulsing barrier electrode B$_N$, an increase in voltage $vB_N$ not only reduces the potential-barrier height between QD$_{N-1}$ and QD$_N$, but also attracts the QDs towards each other, further enhancing their effective coupling~(Fig. \ref{fig:crosstalk}\textbf{d})~\cite{qiao2020coherent}. The orange curve in Fig. \ref{fig:crosstalk}\textbf{d} illustrates this effect for the case of simultaneous pulses applied to electrodes B$_6$ and B$_8$, attracting QD$_5$-QD$_6$ and QD$_7$-QD$_8$ towards each other. This dot movement in turn reduces the exchange coupling between QD$_{N-2}$ and QD$_{N-1}$ as well as QD$_{N}$ and QD$_{N+1}$.

\begin{figure*}[hbt]
	\includegraphics[width=\textwidth]{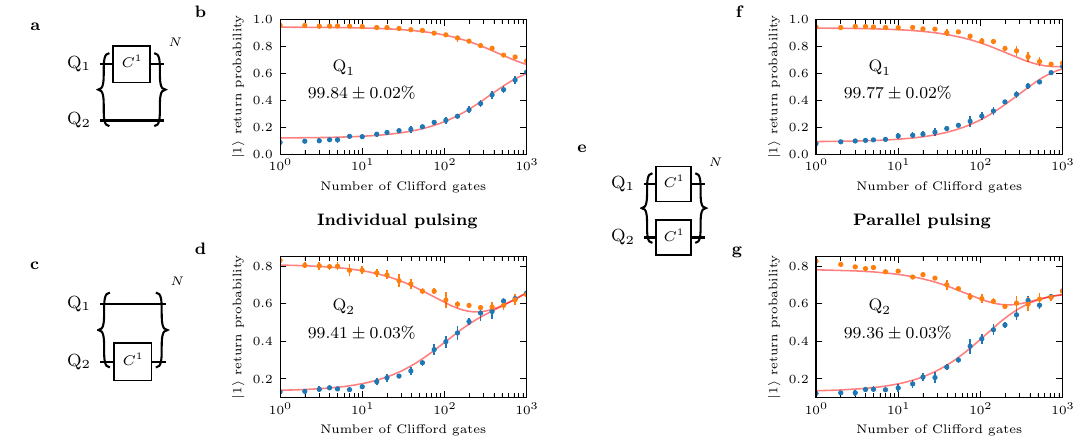}
	\caption{\textbf{Individual vs simultaneous single-qubit gates}~\textbf{a,b)} Blind randomized benchmarking of Q$_1$ with both qubits initialized. We measure an average Clifford fidelity of $99.84 \pm 0.02\%$ including $0.08 \pm 0.02\%$ leakage error. The error bar is calculated as a standard deviation to the mean on 5 measurement repetitions. 
    \textbf{c,d)} Blind RB of Q$_2$, measured similarly to Q$_1$. We measure $99.41 \pm 0.03\%$ average Clifford gate fidelity including $0.13 \pm 0.02\%$ leakage error.
    \textbf{e)} Simultaneous blind RB of Q$_1$ and Q$_2$. The two qubits perform different Clifford sequences with the same number of gates at the same time. We measure $99.77 \pm 0.02\%$ (including $0.11 \pm 0.04\%$ leakage error) and $99.36 \pm 0.03\%$ (including $0.16 \pm 0.03\%$ leakage error) average Clifford gate fidelity for \textbf{f)} Q$_1$ and \textbf{g)} Q$_2$, respectively. This constitutes a $0.05-0.07\%$ reduction in fidelity compared to individual blind RB for both qubits.} 
	\label{fig:1QRB}
\end{figure*}

\subsubsection*{Next-nearest neighbor barrier-barrier compensation}

Intuitively, all exchange couplings, including nearest-neighbor elements, would be compensated in the process of barrier-barrier virtualization~\cite{hsiao2020efficient}. This approach would implement ideal compensation and make the system appear insensitive to crosstalk from nearby electrodes (Fig. \ref{fig:crosstalk}\textbf{d}, black curve). 
In the case of nearest-neighbor barrier-barrier virtualization, one applies a positive voltage pulse to barrier electrodes B$_{N-1}$ and B$_{N+1}$ to counteract any dot movement effects and keep the tunnel coupling to neighboring dots unchanged. 
However, by disallowing shifts in the quantum dot position and removing the enhancement of the tunnel coupling from that effect, one reduces the exchange coupling tunability d$J_N/$d$vB_N$ for all barrier electrodes, which is undesirable.

We note that in both single- and two-exchange-only qubit gates, exchange pulses are never issued on two neighboring barrier electrodes.
As a result, only the crosstalk contributions from next-nearest neighbors need to be managed~(Fig.~\ref{fig:crosstalk}).
In fact, pulsing neighboring barrier electrodes simultaneously would correspond to having an electron being involved in two exchanges at the same time~\cite{qiao2020coherent}, giving rise to different operations, i.e. a $J_x$ rotation for exchange-only qubits~\cite{heinz2024fast}, or Toffoli-like gates for LD qubits~\cite{takeda2022quantum}.
Because there is no situation where we turn on neighboring exchange couplings simultaneously, the substantial impact of an exchange pulse issued on B$_{N-1}$ on the effective neighboring barrier voltage $vB_N$ only results in a small change of the resulting exchange $J_N$, as d$J_N/$d$vB_N$ is small in the regime of low $vB_N$.
We find it acceptable for the tunnel coupling between dots QD$_{N-2}$-QD$_{N-1}$ and QD$_{N}$-QD$_{N+1}$ to change while coupling QD$_{N-1}$ and QD$_{N}$, but we ensure that the couplings between QD$_{N-3}$-QD$_{N-2}$ and QD$_{N+1}$-QD$_{N+2}$ remain unchanged. We illustrate this in the blue curve of  Fig.~\ref{fig:crosstalk}\textbf{d} for simultaneous exchange pulses applied to barrier electrodes B$_6$ and B$_8$. We apply additional barrier pulsing to B$_5$, B$_7$ and B$_9$ to implement the next-nearest compensation and correct for crosstalk during simultanous operation. Detailed description of the barrier-barrier virtualization method can be found in Methods section.

It is worth highlighting the regularity displayed by the extracted cross-capacitance matrix (Extended Data Table \ref{ED:matrix}), where we observe repeated elements for plunger-plunger, plunger-barrier and barrier-barrier elements. This regularity indicates a degree of lateral translational symmetry, confirming good sample uniformity and the benefits of the industrial manufacturing processes \cite{neyens2024probing,george202412}. The observed uniformity reduces the calibration time necessary for virtualization and allows values from a small section (or simulations) to be used as initial guess for the rest of the cross-capacitance matrix or to be directly transferred to lithographically identical devices.

\begin{figure*}[!hbt]
	\includegraphics[width=\textwidth]{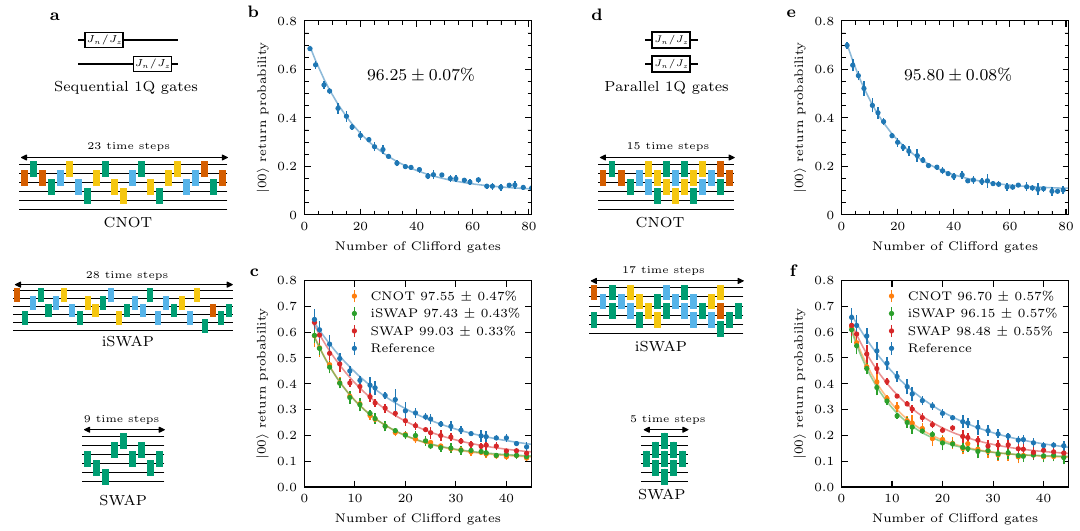}
	\caption{\textbf{Sequential vs parallelized two-qubit gates}~\textbf{a)} Components of sequential two-qubit Clifford gate sequences. We execute single-qubit gates sequentially, and two-qubit gates are executed as sequential exchange pulses. Colors indicate exchange pulses with a $\pi$-rotation (green, spin-swap), a $\frac{3}{2}\pi$-rotation (yellow) and a $\frac{1}{2}\pi$-rotations (blue). Orange indicates custom rotation angles necessary for specific gate implementations (CNOT $\rightarrow$ 2.007~rad, 4.689~rad, 4.736~rad, 1.134~rad; iSWAP $\rightarrow$ 0.162~rad, 4.550~rad).
    \textbf{b)} Two-qubit randomized benchmarking yields a Clifford gate fidelity of $96.25 \pm 0.07\%$. 
    \textbf{c)} Interleaved RB of CNOT, iSWAP and SWAP gates and standard two-qubit RB as reference (measured at the same time as interleaved RB for best accuracy). We measure the fidelity of the CNOT gate to be $97.55 \pm 0.47\%$, iSWAP $97.43 \pm 0.43\%$  and SWAP $99.03 \pm 0.33\%$. 
    \textbf{d)} Components of two-qubit Clifford sequences with parallel pulsing. We apply single-qubit gates simultaneously to both qubits and maximally parallelize exchange pulses in two-qubit gates. 
    \textbf{e)} Two-qubit RB using parallel pulsing. We measure a two-qubit Clifford fidelity of $95.80 \pm 0.08\%$. 
    \textbf{f)} Parallel interleaved RB of CNOT, iSWAP and SWAP gates and standard two-qubit RB as reference. We measure that the fidelity of the parallelized CNOT gate is $96.70 \pm 0.57\%$, parallelized iSWAP is $96.15 \pm 0.57\%$ and parallelized SWAP is $98.48 \pm 0.55\%$. The error bars are estimated as a standard deviation on the mean value of 5 measurement repetitions.} 
	\label{fig:2QRB}
\end{figure*}

\subsubsection*{Individual vs simultaneous single-qubit operation}

We measure the individual qubit performance utilizing the blind RB protocol~\cite{andrews2019quantifying} (see Methods).
To avoid changes in the control scheme, both qubits are initialized, but only one qubit is operated at any time.
We apply a random sequence of $N$ Clifford gates, including a final recovery gate that returns the qubit state to either $\ket{0}$ or $\ket{1}$. 
The return probability decay as a function of increasing sequence length is fitted, and Clifford gate fidelities of $99.84 \pm 0.02\%$ (including $0.08 \pm 0.02\%$ leakage error) for Q$_1$ (Fig.~\ref{fig:1QRB}\textbf{b}) and $99.41 \pm 0.03\%$ (including $0.13 \pm 0.02\%$ leakage error) for Q$_2$ (Fig.~\ref{fig:1QRB}\textbf{d}) are extracted.

Next, we examine the simultaneously-driven single-qubit performance.
We generate multiple sequences of independent, random Clifford gates with increasing length for the individual qubits. 
We initialize both qubits to their $\ket{00}$ state and then execute the respective sequences simultaneously. As single-qubit Clifford gates vary in length, it is possible that one qubit finishes manipulation earlier. It would then idle until the other qubit completes its random Clifford sequence.
As each of the qubits consists of 3 quantum dots and two exchange axes, simultaneous pulses will be issued on the various combinations of exchange axes: $J_z$-$J_z$ (most distant), $J_z$-$J_n$, $J_n$-$J_z$, $J_n$-$J_n$ (next-nearest gates) or individually $J_n$ or $J_z$ on qubits Q$_1$-Q$_2$. 
This probes the average single qubit performance in the presence of simultaneous pulsing. 
We measure that the Clifford gate fidelity for Q$_1$ is~$99.77 \pm 0.02\%$ (including $0.11 \pm 0.04\%$ leakage error) and for Q$_2$ is~$99.36 \pm 0.03\%$ (including $0.16 \pm 0.03\%$ leakage error)(Fig.~\ref{fig:1QRB}\textbf{f, g}), constituting a $0.05-0.07\%$ fidelity reduction compared to individual pulsing. 
This is an encouraging result as it confirms that our techniques limit the impact of crosstalk and enable simultaneous control of neighboring exchange-only qubits, an important functionality for efficient quantum computation.

\subsubsection*{Two-qubit gates with simultaneous exchange pulses}

Parallelizing exchange pulses not only benefits single-qubit control of exchange-only qubits, but has also important implications for the two-qubit gate design. 
A longstanding issue for exchange-only qubits is the complexity of the two-qubit gates, which consist of substantially longer exchange pulse sequences than the single-qubit gate sequences. 
This implies that the two-qubit gate duration, being significant in comparison to typical coherence times, limits the operation fidelity. 
We note that the exchange pulse sequences for CNOT, iSWAP and SWAP gates contain many commuting exchange pulses. 
If exchange pulses commute, they can be executed at the same time, thereby reducing the 23 pulse-long CNOT sequence to 15 time steps, the 28 pulse-long iSWAP sequence to 17 time steps and 9 pulse-long SWAP sequence to 5 time steps. 
This constitutes up to a $\approx 40\%$ reduction in the two-qubit gate duration. Shorter two-qubit gates are beneficial, as they reduce error originating from decoherence. This is true for both qubits involved in the two-qubit gate, but also reduces the errors accumulated by idle qubits waiting for the two-qubit gate completion.

We characterize the performance of our two-exchange-only-qubit system with two-qubit RB. 
As with single-qubit RB, we execute random gate sequences of length $N-1$ using sequential exchange pulses, followed by a recovery gate to $\ket{00}$. 
The gates in the random sequence are sampled from the two-qubit Clifford gate set, containing 11,520 different operations (see Methods). 
Next to the SWAP and CNOT two-qubit gates, we make explicit use of the available iSWAP gate in the two-qubit Clifford gate decomposition, bringing the average number of exchange pulses to 32.3 per Clifford gate. 
We also leverage \textit{mirroring gates} to adjust the configuration of the two exchange-only qubits to shorten two-qubit gate sequences and to issue the majority of exchange pulses on an exchange axis with a high quality factor (see Methods).
By fitting the decay of the $\ket{00}$ recovery probability, we extract a two-qubit Clifford gate fidelity of $96.25 \pm 0.07\%$ (Fig.~\ref{fig:2QRB}\textbf{a,b}). 
We note that this version of two-qubit RB could be improved to better account for leakage (see Extended Data Fig. \ref{fig:full2QRB}; similar to blind single-qubit RB), as the decay to each of the two-qubit basis states is different (see Methods). 

\begin{figure}[!tb]
	\includegraphics[width=\columnwidth]{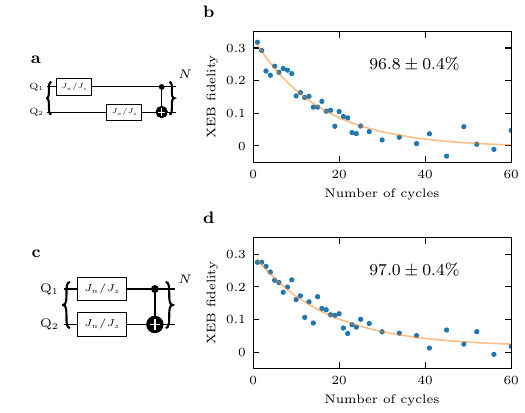}
	\caption{\textbf{Two-qubit cross-entropy benchmarking}~\textbf{a)} Experimental implementation using sequential exchange pulsing. For single-qubit gates, we use randomly selected $J_n$ and $J_z$ axis rotations. We measure a per-cycle fidelity of $96.8 \pm 0.4\%$. 
    \textbf{b)} Experimental implementation using parallel exchange pulsing. The $J_n$ and $J_z$ rotations are issued in parallel, and the maximally parallel version of the CNOT gate is used. We measure a per-cycle fidelity of $97.0 \pm 0.4\%$.}
	\label{fig:xeb}
\end{figure}

In order to characterize the two-qubit gates themselves, we implement interleaved randomized benchmarking (IRB).
Here, the gate to be characterized, e.g. the CNOT gate, is inserted between all the two-qubit Clifford gates of a random gate sequence~\cite{kelly2014optimal}. 
The recovery gate is calculated by taking into account the interleaved gate.
We fit the decay of the $\ket{00}$ recovery probability for the  reference curve (Fig.~\ref{fig:2QRB}\textbf{c}, blue) and compare it to the recovery probability decay when gates are interleaved (Fig.~\ref{fig:2QRB}\textbf{c}, orange - CNOT, red - SWAP, green - iSWAP). 
We measure that the fidelity of the CNOT gate is $97.55 \pm 0.47\%$, iSWAP is $97.43 \pm 0.43\%$  and SWAP is $99.03 \pm 0.33\%$.
We note the trend of increased error per gate with increased number of exchange pulses per gate.

We perform Monte Carlo simulations (see Methods) to evaluate impact of magnetic noise and charge noise on our two-qubit gates. We estimate $T_2^*$ to contribute with $\approx 1.03\%$ to the CNOT gate error (SWAP $\approx 0.17\%$, iSWAP $\approx 1.22\%$), and $\approx 0.48\%$ is due to charge noise (SWAP $\approx 0.14\%$, iSWAP $\approx 0.56\%$). 
The remaining error of $\approx 0.94\%$ is composed of calibration errors (including $J(V_B)$ deviating from the fitted function), non-Markovian errors (such as pulse overlaps) and other unaccounted-for effects (SWAP $\approx 0.66\%$, iSWAP $\approx 0.79\%$). 

Next, we implement two-qubit gates with parallel exchange pulses. 
The two-qubit Clifford gates are performed by executing the single-qubit gates on both qubits in parallel (see Methods), and we utilize the reduced-length two-qubit gates with up to 3 exchange pulses at the same time, as shown in Fig.~\ref{fig:2QRB}\textbf{d}.  
With this parallelization, the average number of exchange pulse time steps is 20.3 per Clifford gate, a reduction of~$\approx$~37~\% in comparison to the sequential pulse execution.
Using two-qubit RB, we measure an average fidelity of $95.80 \pm 0.08\%$ per Clifford gate (Fig.~\ref{fig:2QRB}\textbf{e}). 
We then use IRB to extract the fidelities of the parallelized gates to be $96.70 \pm 0.57\%$ for parallelized CNOT, $96.15 \pm 0.57\%$ for parallelized iSWAP and $98.48 \pm 0.55\%$ for parallelized SWAP. 
For the parallelized CNOT gate, we estimate error contributions of $\approx 0.48\%$ from nuclear spins (SWAP $\approx 0.05\%$, iSWAP $\approx 0.43\%$), and almost unchanged contributions from charge noise of $\approx 0.45\%$ (SWAP $\approx 0.13\%$, iSWAP $\approx 0.56\%$). 
Other effects such as non-Markovian contributions and calibration errors account for $\approx 2.37\%$ (SWAP $\approx 1.34\%$, iSWAP $\approx 2.86\%$).  A comparison of sequential and parallelized two-qubit gate fidelities is provided in an Extended Data Table \ref{ED:comparison_table}. The gate error from nuclear spins typically scales as $\left( t_{gate}/T_2^* \right)^2$, where $t_{gate}$ is the gate time \cite{weinstein2023universal}. We therefore note a substantial decrease by over $60 \%$ in the nuclear spin error contribution for parallelized two-qubit gates, as compared to sequential operation.
While the error contribution from nuclear spins has decreased, the calibration precision of parallel pulses is not sufficient to observe an overall improvement in two-qubit gate fidelity.
This behavior differs from the case of single-qubit gate fidelities, where simultaneous and parallel performance was comparable up to $\approx 0.05-0.07\%$.
We attribute this to a miscalibration of the coupling between the barrier electrode B$_7$ (controlling the exchange between the two qubits) to the barrier electrodes B$_5$ and B$_9$. 
Specifically, we see a change in the fingerprint shape due to simultaneous pulsing that is larger than in other exchange pairs. 
We attribute this to the asymmetric tuning of the dots neighboring the reservoir, where we intentionally reduce the tunnel coupling to the reservoir ($t_c^{\text{res}}$) to isolate the QD array.
Dot-to-dot tunnel coupling is much larger than $t_c^{\text{res}}$. 
This problem can be alleviated with a more consistent tuning, operating on dots inside a larger quantum dot array, or by introducing quadratic components to the crosstalk compensation.

\subsubsection*{Cross-entropy benchmarking}

As a final experiment, we implement cross-entropy benchmarking (XEB)~\cite{arute2019quantum} for both sequential (Fig.~\ref{fig:xeb}\textbf{a}) and parallel exchange pulsing (Fig.~\ref{fig:xeb}\textbf{b}). 
We choose to employ XEB as a characterization tool as it produces an easy-to-track overall performance metric that can be used for systems consisting of up to several dozens of qubits.
Each XEB experiment consists of multiple cycles, which in turn consist of random single-qubit gates or exchange pulses, followed by an entangling two-qubit gate.
When running the parallel version of the experiment, the single-qubit gates are applied simultaneously in each cycle, and the parallel version of the two-qubit gate is issued.
The resulting figure of merit is obtained by comparing the measured bit-string distribution for 100 random circuits against their expected bit-string distribution as determined by simulations~\cite{guerreschi2020intel}.
We measure a per-cycle XEB fidelity of 96.8\% and 97.0\% for the sequential and simultaneous operating modes, respectively, which are matched within the error bars.
We find that the extracted error rate per cycle is comparable to those from two-qubit RB experiments. This is not surprising as the composition of two-qubit Clifford gates closely resembles the base structure of the Random Quantum Circuits (RQC) used in XEB (see Supplementary materials for more information).

\subsubsection*{Conclusions}

The results presented in this paper demonstrate that it is possible to perform high-quality, simultaneous exchange pulses on next-nearest electrodes, allowing for maximum parallelization of quantum circuits in exchange-only qubits. 
An important enabling factor for this experiment is the low defect density of the studied device, which is manufactured on a state-of-the-art \SI{300}{\milli\meter} semiconductor process line.
Still, we find that certain combinations of parallel exchange pulses on the studied device cannot be fully compensated solely by linear virtualization of the corresponding barrier electrodes due to the resulting change in fingerprint shape.
Quadratic corrections to the crosstalk compensation~\cite{rao2024mavis} will mitigate this issue.
We show an experimental implementation of an exchange-only iSWAP gate, and identify a way to schedule parallelized two-qubit Clifford gates.
We use randomized benchmarking to show that the resulting overall quality of parallel exchange pulses is sufficient to match the performance of sequentially issued single-qubit gates.
We execute parallelized two-qubit gates with only a 0.55\% to 1.28\% reduction in fidelity as compared to sequential two-qubit gates, while reducing the gate duration by $\approx 40\%$.
A detailed breakdown of resulting error-budget shows the need for further refinement of pulse control, and highlights the importance of improving material characteristics by utilizing higher-purity isotopes and reducing charge noise.
Furthermore, we adapt cross-entropy benchmarking as a technique to characterize the overall performance of a semiconductor-based quantum system.

\section*{Methods}
\subsection*{Sensor dot readout}
Readout of the exchange-only qubits is performed via Pauli spin blockade \cite{petta2005coherent}, where an electron either tunnels or remains confined depending on its spin state. Nearby sensing dots, that function as single electron transistors (SET), are capacitively coupled to the quantum dot electrons and the SET’s conductance will change depending on whether a charge tunnels or not. In this setup, the change in the SET conductance is detected using a heterojunction bipolar transistor (HBT) based cryogenic amplifier \cite{curry2015cryogenic}. The amplifier and SET are embedded in a lock-in detection circuit where signals of around 50 $\mu \textrm{V}_\textrm{rms}$ at 1 MHz are biasing one ohmic of the SET and the other ohmic is connected to the input of a dual-stage, AC-coupled voltage amplifier. Readout performance in this first-generation cryoamplifier is limited by noise, voltage division, and bandwidth set by the 100 k$\Omega$ shunt resistors.

\subsection*{Device fabrication}
All measurements are performed on an Intel Tunnel Falls device that was fabricated using state-of-the-art high-volume manufacturing techniques.
The QDs are electrostatically confined in a 2DEG defined by a Si/SiGe heterostructure. 
To protect electrons trapped in the QDs from incurring noise due to surrounding nuclear spins, we use $^{28}$Si, isotopically enriched to \SI{800}{ppm}~$^{29}$Si remnants, to grow the quantum well.
Although valley splittings were not measured on this particular device, we extract a range of 50-\SI{140}{\micro eV} for the 1-electron valley splitting on devices from the same \SI{300}{\milli\meter} wafer.
Additional details on device fabrication can be found in ref.~\cite{george202412}, and an extensive discussion regarding device uniformity is presented in ref.~\cite{neyens2024probing}.

\subsection*{Device tuning}
A two dimensional electron gas (2DEG) is accumulated under gate electrodes P$_1$-P$_3$ and P$_{10}$-P$_{12}$ to facilitate access of the middle QDs to an electron reservoir. 
We tune quantum dots QD$_4$-QD$_9$ to the (1,3,1,1,3,1) electron occupation (charge stability diagrams are presented in Extended Data Fig.~\ref{fig:ED_dd}\textbf{a,f,k,p,u}).
The two exchange-only qubits Q$_1$ and Q$_2$ are encoded on dots QD$_4$-QD$_6$ and QD$_7$-QD$_9$, respectively. 
Quantum dots 5 and 8 are populated with three electrons to increase the PSB readout window size that would otherwise be limited by valley splitting~\cite{harvey2018high}; this makes readout more robust against charge drift and miscalibration.
PSB readout is established on dot pairs QD$_4$-QD$_5$ and QD$_8$-QD$_9$ to minimize crosstalk due to capacitive coupling of quantum dots in the QD array. 
Extended Data Fig.~\ref{fig:readout_crosstalk} demonstrates the necessity of a two-lattice-site separation between QD pairs used for PSB in the scenario where both QD pairs are pulsed simultaneously for readout. 
Charge sensing is performed using two sensing quantum dots (SD$_5$ for Q$_1$ and SD$_8$ for Q$_2$) that host many electrons; these are placed on the opposite side of the central screening gate.
With this readout arrangement, we find that the sensitivity of SD$_5$ to QD$_8$-QD$_9$ (performing PSB for Q$_2$), and of SD$_8$ to QD$_4$-QD$_5$ (performing PSB for Q$_1$), is small enough to not be a concern.

\subsection*{Definition of exchange-only qubits}
To define an exchange-only qubit, three spins in three quantum dots are needed~\cite{kempe2001theory, divincenzo2000universal}. 
The computational states of exchange-only qubits are defined as $\ket{0} = \ket{S}\ket{\downarrow}$ ($\ket{0} = \ket{S}\ket{\uparrow}$) and $\ket{1} = \sqrt{\frac{2}{3}}\ket{T_+}\ket{\downarrow} - \sqrt{\frac{1}{3}}\ket{T_0}\ket{\uparrow}$ ( $\ket{1} = \sqrt{\frac{2}{3}}\ket{T_+}\ket{\uparrow} - \sqrt{\frac{1}{3}}\ket{T_0}\ket{\downarrow}$, where $\ket{S} = \sqrt{\frac{1}{2}}(\ket{\uparrow\downarrow} - \ket{\downarrow\uparrow}$, $\ket{T_0} = \sqrt{\frac{1}{2}}(\ket{\uparrow\downarrow} + \ket{\downarrow\uparrow}$ and $\ket{T}_+ = \ket{\uparrow\uparrow}$.
The other possible states are leakage states outside of the computational subspace, which have the same readout signature as the $\ket{1}$ state.
For qubit manipulation, the QD pair with which PSB readout is performed defines the $z$-axis of the qubits' Bloch sphere, and rotations around this $J_z$ axis can be performed by turning on the exchange interaction between these QDs.
The second axis of control is given by $J_n$, positioned at $120^{\circ}$ from the $J_z$ axis, and can be controlled by turning on the exchange interaction between one of the QDs involved in PSB and the gauge spin (third QD).
We initialize the qubits using a charge-locking PSB readout sequence, recording the state of the system prior to qubit manipulation. While this method is not scalable to larger systems as the fraction of accepted records is only $1/4^N$ for $N$ qubits, it allows reducing the setup complexity.
An alternative initialization approach employs deterministic initialization via a reservoir~\cite{botzem2018tuning}. 
As we cannot discern the computational $\ket{1}$~state from leakage states, only records indicating the computational $\ket{0}$~state on both qubits can be used for data analysis. 
While an external magnetic field is not strictly required for exchange only qubit operation, as the gauge spin does not need to be initialized is a specific state, we apply a small 1~mT magnetic field to suppress nuclear spin dynamics \cite{weinstein2023universal}.
For additional detail on the exchange-only qubit encoding, please refer to reference \cite{kempe2001theory, divincenzo2000universal} and see Supplementary materials.

\subsection*{Readout considerations for multi-exchange-only-qubit systems}
Exchange-only qubits are encoded in a decoherence-free (or noiseless) subsystem of three spins in three quantum dots~\cite{kempe2001theory, divincenzo2000universal}.
In the tuning chosen here, the state of Q$_1$ (Q$_2$) can be determined by performing PSB readout on the QD$_4$-QD$_5$ (QD$_8$-QD$_9$) pair.
Neither of the exchange-only qubit computational states are ground- or eigenstates of the three-spin system, which implies that the qubits can be brought into leakage states through initialization, excitation, relaxation or dephasing.
In particular, the computational states will be mixed with the leakage states during idle operations on the timescale of the singlet lifetime~$T_2^*$ (Extended Data Fig. \ref{fig:ED_dd}).
Experimentally, this implies that idle time, even for the initialized $\ket{0}$ state, needs to be avoided. 
Similarly, any idle time introduced before readout will lead to information loss on the $T_2^*$ timescale, rather than the single-spin relaxation time scale $T_1$ relevant for LD qubits. 
In turn, this means that we cannot simply perform sequential readout of both qubits as the second qubit would decay, unless (i), the readout integration time is much shorter than $T_2^*$ (typ. 2-\SI{3}{\micro\second} in \SI{800}{ppm}~$^{28}$Si, but the readout integration time used in this experiment is \SI{18}{\micro\second} per PSB readout) or (ii), we introduce dynamical decoupling to extend the lifetime of the idle qubit~\cite{sun2024full} or (iii), we turn on a sufficiently large exchange such that the PSB pair Hamiltonian eigenbasis corresponds to the qubit readout basis~\cite{zhang2024universal} or (iv), we involve ancillary QDs and swap the PSB QD-pairs away from each other such that both can be read in parallel with individual sensors. 
In our device, the two neighboring sensors share an ohmic contact and part of the accumulated 2DEG, both with substantial series resistance. 
This couples the two SET drain currents, which in turn prohibits the fully parallel readout of neighboring qubits, even though they have their own sensors.
We need to either implement one of the solution listed above or find new technique that would allow us to recover the complete two-qubit information.

\subsection*{Charge-locking readout}
Charge-locking readout is a multi-qubit readout protocol where we first pulse all participating quantum dot pairs to their respective PSB readout window, and only then start signal integration. In this manner, we prevent information loss through decoherence or relaxation while qubits are idling at their respective manipulation points (see Methods). This could be supplemented by the frozen PSB technique~\cite{nurizzo2023complete} if the relaxation times at the PSB window are not significantly longer than the signal integration time (steps III and VI in Fig \ref{fig:device}\textbf{b}).
The steps of the charge-locking PSB readout are described as follows: (I) the first QD pair used for PSB is pulsed into the readout window using a \SI{21.84}{\nano\second} long voltage ramp.
(II) At this readout point, a \SI{3.64}{\nano\second} wait time allows the spins to either be projected to a Singlet and move to the (0,4) charge state, or be projected to a Triplet and remain in the (1,3) charge state.
(III) The tunnel coupling between QD$_4$ and QD$_5$ can be lowered to suppress relaxation rates of these charge states~\cite{nurizzo2023complete}. 
(IV) After projection of the first QD pair, the second QD pair is ramped to its readout window, (V) spins are the projected to Singlet or Triplet states and (VI) the charge relaxation rates are suppressed.
We find the charges to be locked in the readout window, with relaxation times substantially longer than the integration time of \SI{18}{\micro\second} required to obtain a sufficient-quality signal.
Therefore, we deem steps (III) and (VI) not strictly necessary in our experiment.
Once both qubits have been projected, a voltage pulse is applied to the SD$_5$ plunger to shift its electrochemical potential to the flank of a Coulomb peak to allow for charge sensing. 
(VII) Then, a pulsed readout stimulus signal is applied to the sensor that allows determination of the state of Q$_1$.
(VIII) After the readout of Q$_1$ is completed, the SD$_5$ plunger is brought back into Coulomb blockade, and the process is repeated for Q$_2$.
We achieve a signal-to-noise ratio (SNR) of 4.25 and 3.1 for readout of Q$_1$ and Q$_2$, respectively. 
(IX-X) At the end of the sequence, both qubits are pulsed back to their manipulation point.

\subsection*{Virtualization}

In order to efficiently control the array of quantum dots, we employ standard virtualization techniques~\cite{hensgens2017quantum, volk2019loading}. 
The relative effect between the dedicated plunger electrode and the surrounding plunger and barrier electrodes on the chemical potential of a QD are measured by tracking electron addition lines in charge stability diagrams as function of the applied voltages. 
These values are collected in the cross-capacitance matrix, which in turn is inverted to perform a linear basis transformation into the \textit{virtual} voltage space. 
These virtual voltages now only affect the chemical potential of the quantum dot they are supposed to control, leaving the chemical potentials of other dots unaffected.

The voltage virtualization is fine-tuned after the quantum dot array is brought into the correct charge configuration and is isolated from the environment (Extended Data Fig.~\ref{fig:ED_dd}~\textbf{a,f,k,p,u}).
However, at that point, charge addition lines are no longer visible and can't be used to update the voltage virtualization.
Therefore, the crosstalk matrix entries are adjusted in an iterative process where the position of inter-dot charge transition lines is monitored as virtual voltages on surrounding electrodes are changed.
Proper virtualization is achieved once the position of charge transition lines does not change with altering virtual voltages.
This procedure can be used for plunger-plunger as well as for plunger-barrier crosstalk compensation; the technique for barrier-barrier crosstalk compensation is described in the main text.

\subsection*{Barrier-barrier crosstalk compensation}
To compensate the barrier-barrier crosstalk, the center line-cut of~Fig.~\ref{fig:crosstalk}\textbf{a} is tracked as a function of the simultaneous barrier pulse amplitude~$vB_6$ (Fig.~\ref{fig:crosstalk}\textbf{e}). 
As $vB_6$ is increased, we observe an increase of the exchange between QD$_7$ and QD$_8$, in line with our observations in Fig.~\ref{fig:crosstalk}\textbf{c}.
We note that the exchange pattern is linearly shifted as a function of $vB_6$, indicating that the linear crosstalk compensation framework is applicable.
The value of the crosstalk element is given by the slope of the resulting line pattern.
We enter it into the existing voltage virtualization framework that relies solely on a linear basis transformation.
In addition, we include finer corrections to compensate any outstanding movement along the detuning axis by optimizing barrier-plunger elements. 
Furthermore, we note a slight fingerprint shape change that reduces the exchange tunability during simultaneous exchange pulses as compared to individual exchange pulses for some of the QD pairs (most notably B$_5$ and B$_7$). 
We account for this by adding small nearest-neighbor barrier-barrier elements to the cross-capacitance matrix (Extended Data Table~\ref{ED:matrix}).
After the calibration cycle, we check the dependence of exchange strength as a function of the simultaneous barrier pulse amplitude (Fig.~\ref{fig:crosstalk}\textbf{f}) in comparison to the crosstalk-uncompensated situation (Fig.~\ref{fig:crosstalk}\textbf{e}). 
Using the described corrections, we observe that the resulting exchange pattern is independent of the simultaneously applied exchange pulse.

\subsection*{Exchange pulse calibrations}
Exchange-only qubits are operated by issuing well-calibrated voltage pulses to dynamically control the exchange interaction between the relevant QDs~\cite{kempe2001theory, divincenzo2000universal}. 
We choose to keep the durations $t_p$ and $t_b$ of the resulting exchange pulses equal and constant throughout the experiment.
The rotation angle produced by an exchange pulse is given as $\varphi=t_p\cdot J(V_B)$, and we control it with the voltage $V_B$ issued to the electrode controlling the exchange interaction. 
To calibrate each exchange axis, a qubit containing this exchange axis is encoded and prepared in its $\ket{0}$~state, exchange pulses on the axis under question are applied, and the qubit is read out. 
To ensure adequate measurement contrast (e.g., to calibrate the $J_z$ axis), pre-rotations or excitation swaps leveraging previously calibrated axis are applied.
Then, for each axis, the following three steps are performed:
\begin{itemize}
    \item \textbf{Fingerprint calibration $\rightarrow$} The goal of the fingerprint calibration is to ensure that exchange pulses are insensitive to charge-noise in the double-dot detuning direction. Therefore, the exchange strength is measured as a function of barrier and detuning voltages~\cite{reed2016reduced}. Then, the crosstalk compensation between the exchange barrier electrode and involved plunger electrodes is adjusted to keep the sweet spot along the exchange axis at a constant detuning voltage. Extended Data Fig.~\ref{fig:ED_dd}\textbf{b,g,l,q,v} shows fingerprints for exchange axis used in the experiment.
    While this calibration can be performed with a single exchange pulse, we leverage an error amplification scheme by applying 8 repeated exchange pulses to increase the sensitivity of the calibration.
    
    \item \textbf{Initial exchange calibration $\rightarrow$} Next, we measure the exchange coupling magnitude in the function of the barrier gate voltage at the selected detuning voltage. For this, exchange pulses of increasing amplitudes are applied and the $\ket{1}$~state return probability is measured.
    The resulting data can be fit to $b+a\cdot\cos{\left(J(V_B) t_p\right)}$, where $J(V_B)=\alpha\cdot\exp{\left(\gamma V_B\right)}$ and $\gamma$ is the exchange tunability.
    
    \item \textbf{Fine exchange calibration $\rightarrow$} With an initial knowledge of $J(V_B)$, it is possible to issue pulses of a desired rotation angle on this axis. To amplify the calibration error, a train of $N$ pulses with a fixed rotation angle is measured. As the angle is swept, the resulting oscillations can be used to extract a more precise relation of $J(V_B)$, which can be described as $J(V_B)=\alpha\cdot\exp{\left(\gamma V_B + \kappa V_B^2\right)}$, where $\kappa$ is a second-order correction to the exchange tunability. Additional precision in the calibration can be gained by directly using a heuristic interpolation of $J(V_B)$. The final calibration is presented in the Extended Data Fig.~\ref{fig:ED_dd}\textbf{c,h,m,r,w}.
\end{itemize}

Once the relation $J(V_B)$ is established, pulses with rotation angle of $\varphi$ can be produced by issuing barrier pulses with amplitude

\begin{equation*}
    V_B(\varphi) = \frac{-\gamma +\sqrt{\gamma^2 + 4\kappa\cdot \ln \left( \varphi/(\alpha \cdot t_p)\right)}}{2\kappa} \text{ for }\kappa\neq0,
\end{equation*}
and 
\begin{equation*}
    V_B(\varphi) = \frac{1}{\gamma}\ln\left(\frac{\varphi}{t_p\cdot\alpha}\right)\text{ for }\kappa=0.
\end{equation*}

Furthermore, we ensure negligible error contributions from residual coupling $J_{\text{off}}=J(V_{B_{\text{off}}})$ by reducing the voltage $V_{B_{\text{off}}}$ on the electrode when the exchange coupling is not active.
We need to ensure that this condition is met and that rotation angles of up to $2\pi$ are available within the dynamic range of the Arbitrary Waveform Generator (AWG).
We achieve this by pulsing negatively on the barrier while its coupling is not active to make full use of the available dynamic range of the AWG.

\subsection*{Pulse pre-distortions}
To minimize the effects of decoherence caused by magnetic noise from the spin bath, it is beneficial to operate with as-fast-as practical exchange pulses. 
However, as pulse time $t_p$ and buffer time $t_b$ are reduced, the qubit fidelity is limited by non-Markovian control errors resulting from overlap of subsequent pulses due to filtering effects of the transmission lines used to deliver the pulses.
More concretely, the finite bandwidth of both the control electronics as well as the transmission lines results in the ideally square pulses produced by the control electronics being distorted. 
While a systematic distortion that only affects the current pulse can be accounted for by standard calibrations, finite pulse fall times of one pulse that affect the amplitude of the subsequent pulse create an effective system memory.
This system memory cannot be accounted for by only using standard calibrations. 
Additional effects leading to a system memory can include pulse reflections at impedance mismatches in the signal chain, or the high-pass effect of the bias tees used.
To counter these effects, we implement exponential undershoot/overshoot pulse pre-distortions.
Properly pre-distorted pulses avoid pulse overlap and reduce the pulse rise and fall time~\cite{langford2017experimentally, rol2020time}.

We tackle the task of finding the appropriate pre-distortions in two steps.
First, we introduce a buffer time spectroscopy measurement where $t_b$ is varied for a train of pulses with varying amplitudes (Extended Data Fig.~\ref{fig:buffer_spec}).
Tracking the voltage at which a particular overall rotation angle is achieved as a function of $t_b$ gives a good approximation to the step response. Based on this dependence, exponential parameters are extracted to define the pre-distortion coefficients for short ($\approx$\SI{8}{\nano\second}) and medium ($\approx$\SI{50}{\nano\second}) timescales.

The pre-distortion corrections are applied in real time at the last output stage of the AWG.
The effect of predistorted pulses can be assessed by taking another buffer time spectroscopy measurement where now the approximate step response for these timescales is substantially flatter. 
However, these corrections still leave long pulse sequences of several thousand exchange pulses vulnerable to effects on longer timescales.

To further reduce the effect of non-Markovian errors on longer timescales, a second step to optimize the best pre-distortion parameters is performed. 
As the cost function of the optimization procedure, we choose the visibility of a sequence of random Clifford gates (difference between the recovered $\ket{1}$ and $\ket{0}$ states). 
To reduce the complexity of the procedure, the same pre-distortion parameters are applied on all control lines. 
We find that an exponential model with a timescale of $\approx$\SI{500}{\nano\second} describes the filter function of the transmission lines well. 
We choose to apply these corrections pulse-by-pulse by modifying the amplitude of the exchange pulses based on the history of exchange pulses on the respective channel rather than applying them sample-by-sample.
This is why Extended Data Fig.~\ref{fig:Nosc_vs_Q} can show a time-dependent oscillation frequency, but we can still measure high-fidelity exchange pulses over extended periods of time.
Otherwise, this technique follows the same mathematical framework. 
This procedure can be further improved by designing a targeted error-amplifying sequence to act as the cost-function for the optimization.

\subsection*{Exchange-only gate library}
\label{sec:gate_library}
General single-qubit gates for exchange-only qubits need to be synthesized from positive-angle rotations around the $n$-~and $z$-axes~\cite{kempe2001theory, divincenzo2000universal}.
The precise decomposition of Clifford gates to $n$-~and $z$-exchange pulses can be found in ref.~\cite{andrews2019quantifying}.
With this decomposition, a single-qubit Clifford gate needs on average 2.666~exchange pulses.

Two-qubit gates for exchange-only qubits are also composed solely of exchange pulses. 
However, the precise pulse sequence, and therefore also the number of pulses and the duration of the gate, depend on the connectivity between the qubits as well as their configuration.
The two-qubit gate library implementation of ref.~\cite{chadwick2024short} is used to determine the appropriate sequence of pulses.
The two-qubit gates supported in this library are CNOT, iSWAP and SWAP.
Of the 11,520 two-qubit Clifford gates, 576 contain no two-qubit gate, 576 include a SWAP, 5,184 include a CNOT and 5,184 include an iSWAP gate, with an average of 32.3~exchange pulses.

To improve the efficiency of pulse execution on two qubits, exchange pulses can be applied in parallel.
This can be done straightforwardly in the case of single-qubit gates on the respective qubits.
For two-qubit gates, an algorithm that issues pulses ``as soon as possible'' for the two-qubit gate sequence while avoiding scheduling pulses on neighboring barriers is used.
A general two-qubit Clifford gate can be decomposed into single-qubit Clifford gates acting on either qubit, followed by a two-qubit gate, followed again by single-qubit Clifford gates acting on either qubit.
We implement the parallelization of a general two-qubit Clifford gate by parallelizing its respective parts.

\subsection*{Mirroring gates for exchange-only qubits}
The flexibility of exchange-only qubits can further be enhanced by the use of \textit{mirroring gates}.
Which exchange axis is responsible for $n$-, and which axis is responsible for $z$-rotations - the configuration of the exchange-only qubit - is defined by the location of the QD pair that performs PSB readout.
This gives each exchange-only qubit two possible configurations.
The configuration can be changed by choosing the other possible QD pair to perform PSB readout, or it can be changed in-situ, by applying \textit{mirroring gates} consisting of 3 exchange $\pi$-pulses around the $n-$, $z-$ and $n-$axis.
Here, changing the QD pairs performing readout isn't possible, but applying mirroring gates after initialization and before readout provide us with this additional flexibility.
Choosing the optimal combination of configurations for the two qubits has two advantages.
First, two-qubit gates with fewer pulses can become available~\cite{chadwick2024short}, reducing overall two-qubit gate complexity.
Second, because gates have an unequal distribution of pulses on the various exchange axes, fewer pulses can be scheduled on exchange axes with reduced quality factors, increasing average gate fidelity.

\subsection*{Randomized benchmarking}
We choose to utilize RB as the primary tool to characterize the fidelities of single- and two-qubit gates.
For RB, a random sequence of $N$ single- or two-qubit Clifford gates is applied, followed by a recovery gate that returns the qubit(s) to a known state. 
The Clifford gates are decomposed into exchange pulses as described in the previous section.
By interpolating the state recovery probability as function of $N$, the average error per Clifford gate can be extracted.
In more detail, for the single-qubit experiments, we use a technique called \textit{blind} randomized benchmarking~\cite{andrews2019quantifying} where each seed is run twice, recovered one time to $\ket{0}$ and another time to $\ket{1}$.
Analysis of the resulting curves allows extraction of the leakage error per Clifford, which we find to be in line with expectations given pulse durations and abundance of nuclear spins (see main text).

When performing parallelized RB, single-qubit gates are issued at the same time on the respective qubits, and the parallelized versions of the two-qubit gates in the gate library are used.
We also emphasize that the full 2-qubit state readout is performed.
This ensures that state leakage is properly accounted for, which can be challenging when only partial information of the two-qubit state is extracted (Extended Data Fig. \ref{fig:full2QRB}).
For single- and two-qubit RB, we report the mean fidelity and error of the mean of 5 measurements that each are averaged over 20 random seeds. 
For interleaved two-qubit RB, we use 5 measurements with 25 random seeds each.

\subsection*{Monte Carlo simulation}
To improve our understanding of the system, we estimate the impact of magnetic noise and charge noise with a Monte Carlo simulation by assuming quasi-static noise $\delta B$ and $\delta J$ on both the Zeeman and exchange coupling, respectively.
The two-qubit gate fidelity is then calculated as the average projection of the expected state with that of the final state for all computational states. 
The magnitude of $\delta B$ is chosen to fit the average pure dephasing time $\bar{T_2^*}$, calculated as $\dfrac{1}{\bar{T_2^*}} = \sqrt{\sum_n^N \left(\dfrac{1}{{^nT_2^*}}\right)^2}/N$ of the $N$ QD pairs involved, where $^nT_2^*$ is the singlet lifetime in the $n^{th}$ double dot (Extended Data Fig.~\ref{fig:ED_dd}\textbf{e,j,o,t,y}). 
An appropriate $\delta J$ is chosen to fit the number of resolvable exchange oscillations $N_{osc}$ for each of the exchange axes (Extended Data Fig.~\ref{fig:ED_dd}\textbf{d,i,n,s,x}).
We use $N_{osc}$ rather than a single-frequency quality factor $Q$ of the exchange oscillations, as the experimental data shows a change in the exchange oscillation frequency due to slow settling of the voltage on the gate electrode (Extended Data Fig.~\ref{fig:Nosc_vs_Q}).
We note that the evolution of the exchange coupling remains coherent well past the decay of the fixed frequency fit. 
The parameter $N_{osc}$ is extracted by fitting a Gaussian decay to the oscillation envelope, providing a better estimate for the impact of charge noise as it is insensitive to small changes in the exchange oscillation frequency~\cite{weinstein2023universal}.

\section*{Data availability}
The raw data and analysis that support the findings of this study are available in the Zenodo repository (\href{http://zenodo.org/doi/10.5281/zenodo.15090726}{http://zenodo.org/doi/10.5281/zenodo.15090726}).


\section*{Author contributions}
M.T.M. performed measurements, with assistance from F.L., F.B., J.Z., M.C. and B.H. 
J.D.C., G.G.G., F.L. and M.T.M. developed the gate library. 
F.L. wrote the measurement software, with assistance from J.Z., F.R., F.B., M.T.M. T.M.M. and M.R.
F.A.M. estimated the error contributions. 
G.G.G., F.A.M., S.Pr., A.R. and A.Y.M. developed the XEB implementation for exchange-only qubits. 
S.N., O.K.Z., A.N., P.L.B., J.R., D.K., T.F.W., E.J.C., J.C., R.S., R.O., B.P. and E.E provided feedback used for the development of the fabrication process. 
F.L., F.B., N.K., R.F., R.W.M., L.P.O.I. and S.Pe. developed the pre-distortion framework. 
L.F.L. designed the thermalization hardware. 
H.C., E.H., M.M.I., S.A. and R.P. fabricated the sample. 
M.T.M. and F.L. wrote the manuscript with contributions from F.A.M., G.G.G., N.B., J.R., J.S.C. and input from all authors.

\section*{Competing interests}
The authors declare no competing interests.

\newpage
\onecolumn
\section*{Extended Data figures}
\renewcommand{\figurename}{Extended Data Fig.}
\setcounter{figure}{0}   

\begin{figure*}[!hbt]
	\includegraphics[width=\textwidth]{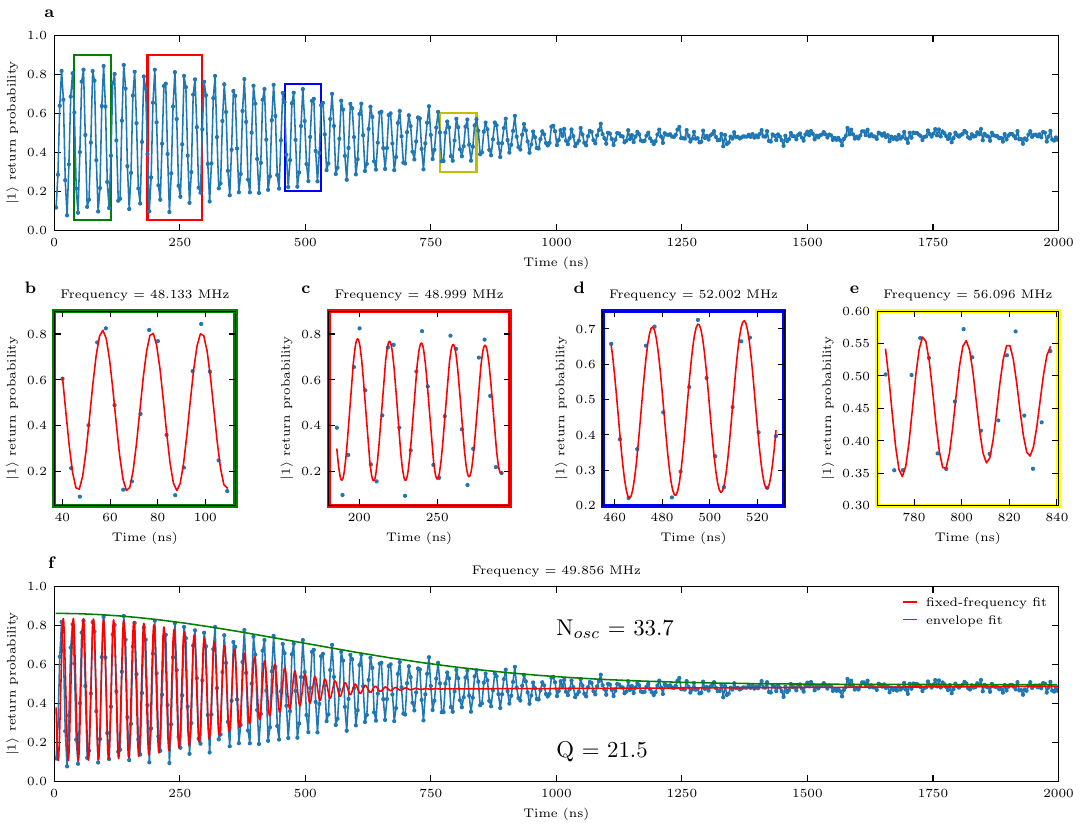}
	\caption{\textbf{Exchange quality factor vs $N_{osc}$}~\textbf{a)} Exchange oscillations measured for QD$_8$-QD$_9$. \textbf{b-e)} Zoom-in on the exchange oscillations with sinusoidal fits to extract the momentary frequency. We note that the frequency increases with increasing pulse length, which can be correlated with the final settling of the barrier electrode voltage. For qubit control, only the total angular evolution during the exchange pulse is relevant. Overlap with transients from previously issued pulses can be accounted for with additional calibrations such as pulse-domain pre-distortions. It is, however, difficult to reliably extract quality factors for exchange oscillations at a fixed frequency. \textbf{f)} When fitting the data with a decaying sinusoid, we extract a quality factor of 21.5 at \SI{49.856}{\mega\hertz}. Ignoring the steadily rising frequency, we also fit the envelope of the decaying oscillation to extract $N_{osc}$ of 33.7~\cite{weinstein2023universal}.} 
	\label{fig:Nosc_vs_Q}
\end{figure*}

\begin{figure*}[p]
	\includegraphics[width=\textwidth]{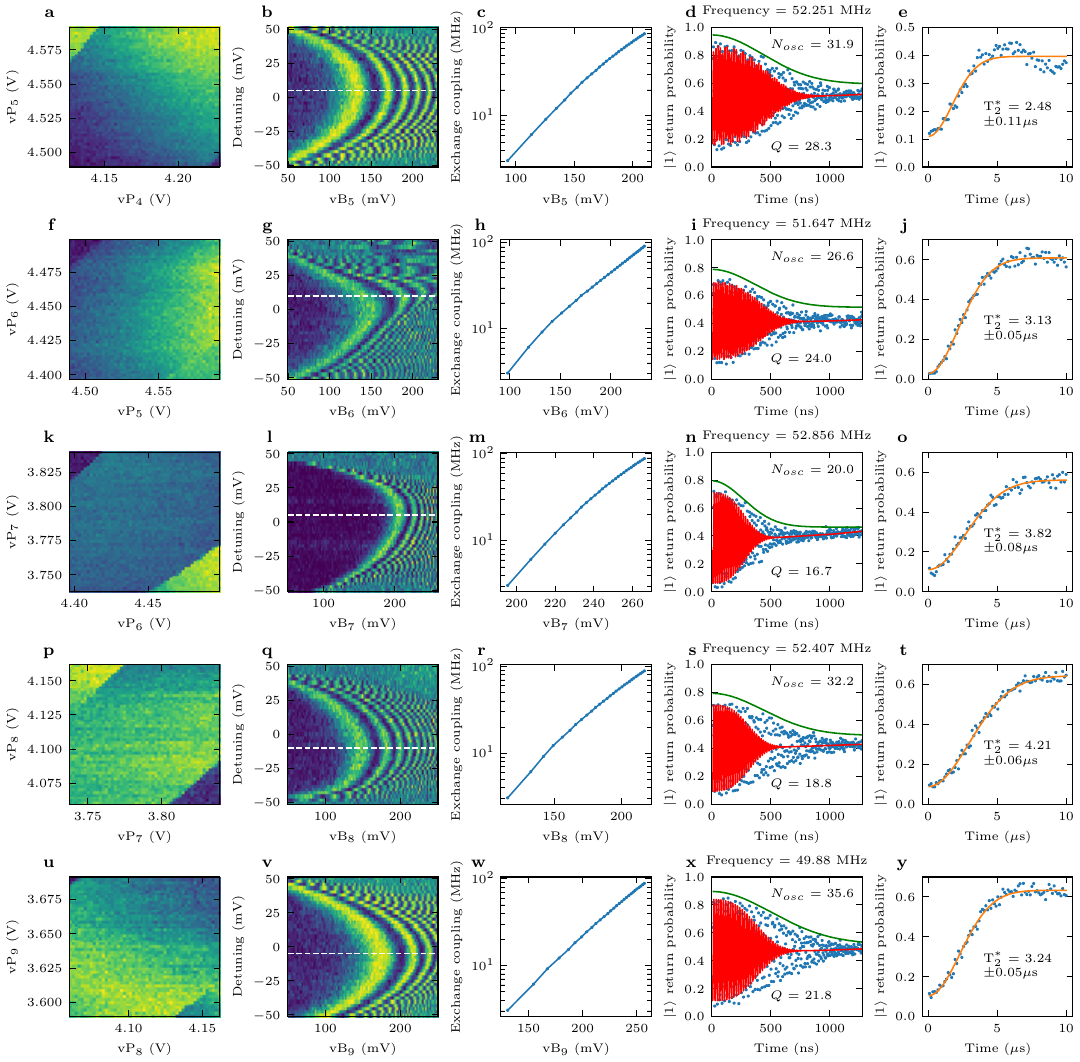}
	\caption{\textbf{Double quantum-dot characterization}~\textbf{a,f,k,p,u)} Charge stability diagrams for \textbf{a)} QD$_4$-QD$_5$ in (1,3) charge configuration, \textbf{f)} for QD$_5$-QD$_6$ in (3,1), \textbf{k)} for QD$_6$-QD$_7$ in (1,1), \textbf{p)} for QD$_7$-QD$_8$ in (1,3), and \textbf{u)} for QD$_8$-QD$_9$ in (3,1). \textbf{b,g,l,q,v)} Exchange fingerprints for \textbf{b)} QD$_4$-QD$_5$ with four \SI{10.92}{\nano\second} pulses, \textbf{g)} for QD$_5$-QD$_6$ with four \SI{10.92}{\nano\second} pulses, \textbf{l)} for QD$_6$-QD$_7$ with eight \SI{10.92}{\nano\second} pulses, \textbf{q)} for QD$_7$-QD$_8$ with eight \SI{10.92}{\nano\second} pulses, and \textbf{v)} for QD$_8$-QD$_9$ with eight \SI{10.92}{\nano\second} pulses. \textbf{c,h,m,r,w)} Calibrated exchange coupling tunability for \textbf{c)}~QD$_4$-QD$_5$, \textbf{h)}~for QD$_5$-QD$_6$, \textbf{m)}~for QD$_6$-QD$_7$, \textbf{r)}~for QD$_7$-QD$_8$, and \textbf{w)}~for QD$_8$-QD$_9$. \textbf{d,i,n,s,x)} Measurement of exchange oscillations for \textbf{d)}~QD$_4$-QD$_5$, \textbf{i)}~for QD$_5$-QD$_6$, \textbf{n)}~for QD$_6$-QD$_7$, \textbf{s)}~for QD$_7$-QD$_8$, and \textbf{x)}~QD$_8$-QD$_9$. \textbf{e,j,o,t,y)}~Singlet lifetime measurement for \textbf{d)}~QD$_4$-QD$_5$,  \textbf{i)}~for QD$_5$-QD$_6$, \textbf{n)}~for QD$_6$-QD$_7$, \textbf{s)}~for QD$_7$-QD$_8$, and \textbf{x)}~for QD$_8$-QD$_9$.} 
	\label{fig:ED_dd}
\end{figure*}

\begin{figure*}[p]
	\includegraphics[width=\textwidth]{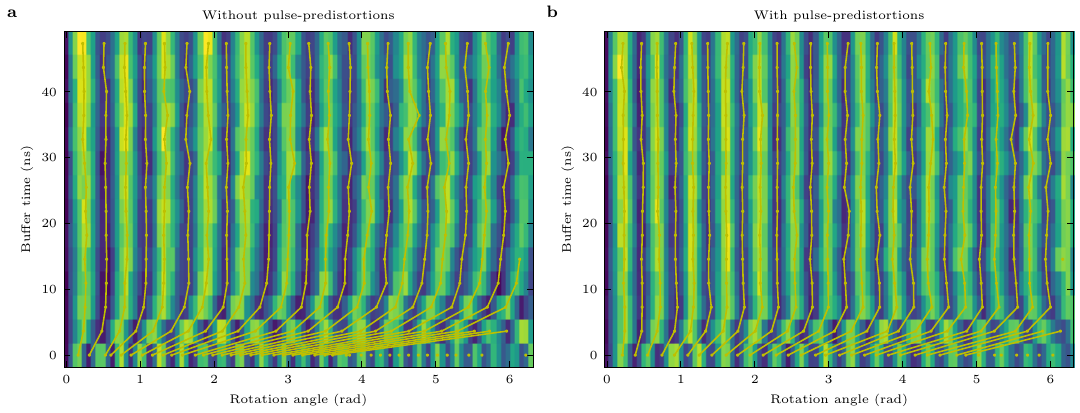}
	\caption{\textbf{Buffer time spectroscopy}~\textbf{a)} We apply a train of 12 consecutive exchange pulses. On the x-axis, we vary the targeted angular evolution for each exchange pulse between 0 and 2$\pi$. On the y-axis, we vary the buffer time $t_b$ between the exchange pulses. Ideally, we expect to see oscillations with 12 periods, one per pulse, regardless of the buffer time. However, in the presence of finite rise and fall times we observe more oscillations at short buffer times, indicating that the overlap of consecutive pulses invalidates our calibrations. \textbf{b)} Buffer spectroscopy measurement after introducing pulse pre-distortions that reduce pulse overlap. The plot is visibly more consistent with expectations. Therefore, buffer times as short as 7.28~ns can be used in the experiment.} 
	\label{fig:buffer_spec}
    
\end{figure*}

\begin{figure*}[p]
	\includegraphics[width=\textwidth]{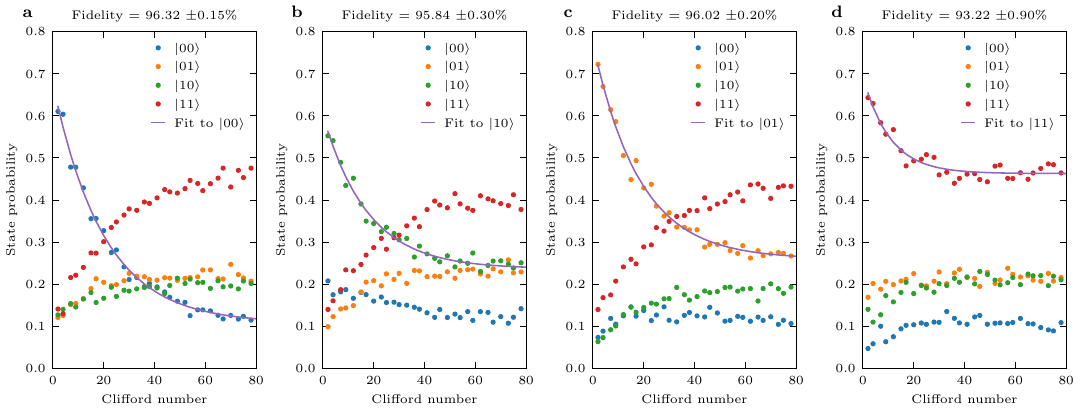}
	\caption{\textbf{Two-qubit RB with recovery to all computational basis states.}~ \textbf{a)} $\ket{00}$, \textbf{b)} $\ket{10}$, \textbf{c)} $\ket{01}$, \textbf{d)} $\ket{11}$ state. The return probability for all two-qubit basis states is recorded. This data can be used for the development of blind two-qubit randomized benchmarking.}
	\label{fig:full2QRB}
\end{figure*}

\begin{figure*}[p]
	\includegraphics[width=\textwidth]{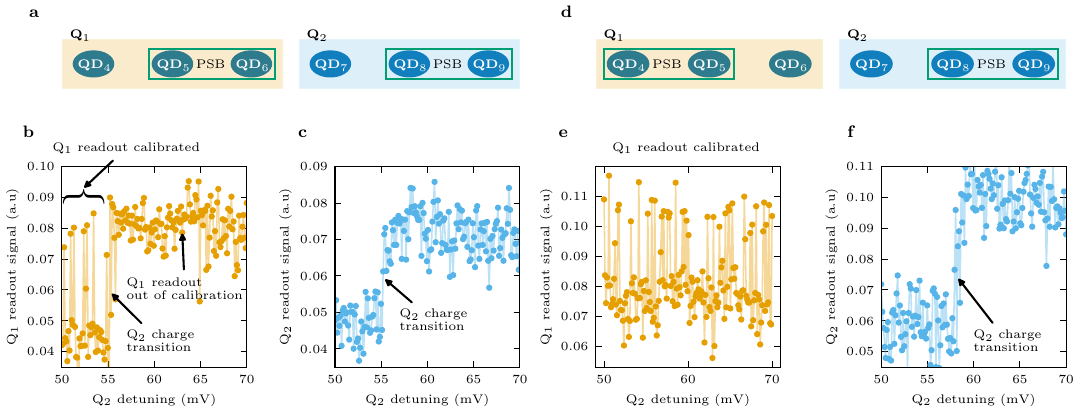}
	\caption{\textbf{Crosstalk in charge-locking PSB readout.}~\textbf{a)} Two-exchange-only-qubit arrangement where PSB readout is performed using QD$_5$-QD$_6$ for Q$_1$ and QD$_8$-QD$_9$ for Q$_2$. \textbf{b,c)} We perform charge-locking readout as described in the main text for Q$_1$ and Q$_2$, with the PSB readout arrangement as visualized in panel \textbf{a)}. We scan the Q$_2$ detuning pulse magnitude, while readout parameters for Q$_1$ remain constant. At each Q$_2$ detuning value we record the PSB readout result for both Q$_1$ and Q$_2$. As we do not modify Q$_1$ readout parameters, we expect to  see a variable signal signifying a Singlet for high results, and a Triplet for low results. Instead, we see that at the point where Q$_2$ crosses a charge transition from (3,1) to (4,0) (panel \textbf{c)}), the Q$_1$ readout is pushed out of calibration and into a (4,0) charge state. This is due to capacitive coupling between quantum dots participating in Q$_1$ and Q$_2$ readout. Similar effects have been used for cascade readout in reference \cite{van2021electron}. \textbf{d)} Two-exchange-only-qubit arrangement where PSB readout is performed using QD$_5$-QD$_6$ for Q$_1$ and QD$_8$-QD$_9$ for Q$_2$. \textbf{e,f)} With two dot separation between PSB readout pairs, Q$_1$ readout remains calibrated, regardless of the Q$_2$ detuning scan or Q$_2$ charge transition.} 
	\label{fig:readout_crosstalk}
\end{figure*}

\begin{figure*}[p]
	\includegraphics[width=\textwidth]{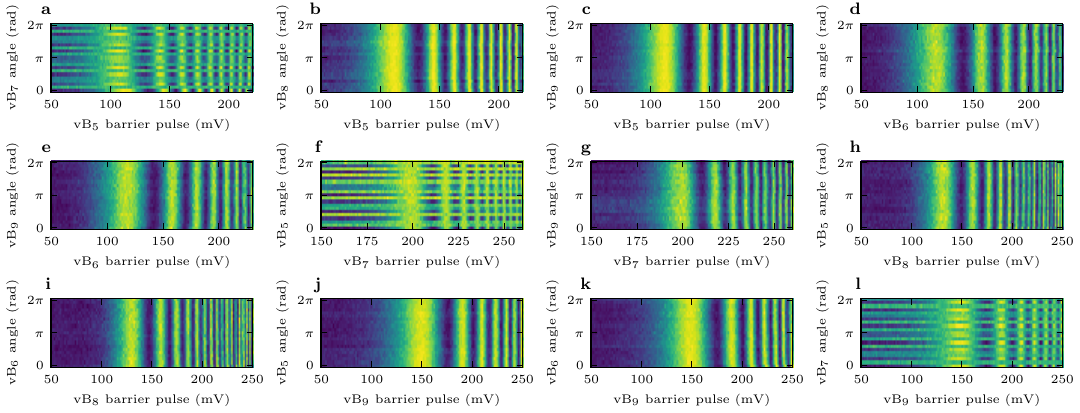}
	\caption{\textbf{Barrier-barrier crosstalk measurements.} Panels \textbf{a,f,l)} show horizontal lines where the total angular evolution under the extra barrier pulse swept on the y-axis leads the exchange-only qubit to leave its computational subspace.} 
	\label{fig:EDcrosstalk}
\end{figure*}

\begin{figure*}[p]
	\includegraphics[width=\textwidth]{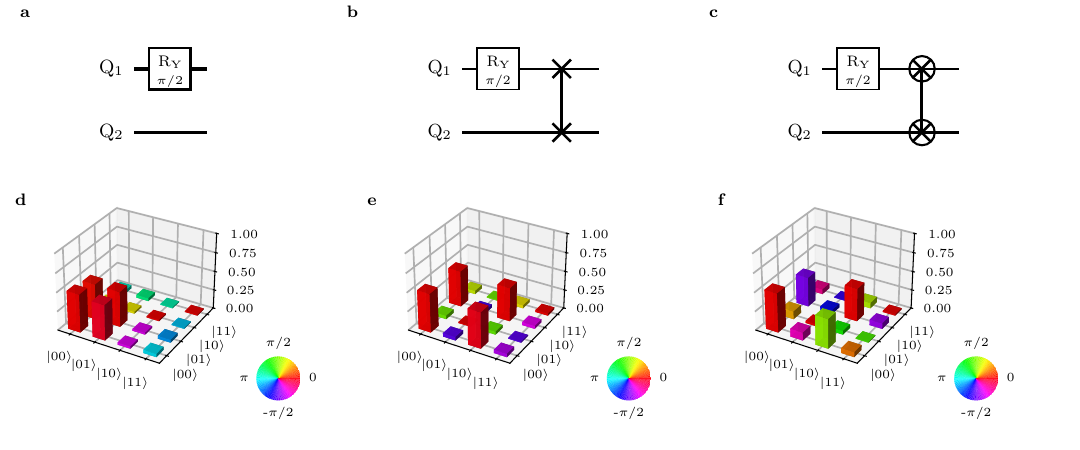}
	\caption{\textbf{Validation of the phase imparted by iSWAP gate}~\textbf{a,b,c)} Quantum circuits executed in this experiment. \textbf{a)} Prepare Q$_1$ in the superposition state $\ket{Q_1Q_2} = \frac{1}{\sqrt{2}}(\ket{0}+\ket{1})\ket{0}$. \textbf{b)} Prepare Q$_1$ in a superposition state and apply a SWAP gate, bringing the system to $\ket{Q_1Q_2} = \ket{0}(\frac{1}{\sqrt{2}}(\ket{0}+\ket{1}))$. \textbf{c)} Prepare Q$_1$ in a superposition state and apply an iSWAP gate, bringing the system to $\ket{Q_1Q_2} = \ket{0}(\frac{1}{\sqrt{2}}\ket{0}+\frac{i}{\sqrt{2}}\ket{1})$ \textbf{d,e,f)} Quantum state tomography results for the circuits in \textbf{a,b,c)}. We note that in addition to swapping the superposition state between qubits, the iSWAP gate imparts the expected $i$ phase. } 
	\label{fig:iswap}
\end{figure*}

\clearpage
\begin{tabular}{|l|rrrrrrrrrrrrrrr|}
\hline
          &   vP$_3$ &   vB$_4$ &   vP$_4$ &   vB$_5$ &   vP$_5$ &   vB$_6$ &   vP$_6$ &   vB$_7$ &   vP$_7$ &   vB$_8$ &   vP$_8$ &   vB$_9$ &   vP$_9$ &   vB$_{10}$ &   vP$_{10}$ \\
\hline
 P$_3$    &     1    &    0.65  &     0.27 &   0.1    &     0.05 &    0     &     0    &   0      &     0    &    0     &     0    &    0     &     0    &       0     &        0    \\
 B$_4$    &     0    &    1     &     0    &   \textbf{0.044}  &     0    &   \textbf{ 0.05}  &     0    &   0      &     0    &    0     &     0    &    0     &     0    &       0     &        0    \\
 P$_4$    &     0.3  &    0.6   &     1    &   0.68   &     0.25 &    0.1   &     0.05 &   0.0025 &     0    &    0     &     0    &    0     &     0    &       0     &        0    \\
 B$_5$    &     0    &    \textbf{0.044} &     0    &   1      &     0    &    \textbf{0.044} &     0    &   \textbf{0.01}   &     0    &    0     &     0    &    0     &     0    &       0     &        0    \\
 P$_5$    &     0.05 &    0.1   &     0.3  &   0.62   &     1    &    0.72  &     0.26 &   0.13   &     0.05 &    0.015 &     0    &    0     &     0    &       0     &        0    \\
 B$_6$    &     0    &    \textbf{0.02}  &     0    &   \textbf{0.044}  &     0    &    1     &     0    &  \textbf{ 0.044 } &     0    &    \textbf{0.018} &     0    &    0     &     0    &       0     &        0    \\
 P$_6$    &     0    &    0.015 &     0.05 &   0.12   &     0.25 &    0.64  &     1    &   0.65   &     0.32 &    0.135 &     0.06 &    0.015 &     0    &       0     &        0    \\
 B$_7$    &     0    &    0     &     0    &  \textbf{ 0.017}  &     0    &    \textbf{0.044} &     0    &   1      &     0    &    \textbf{0.044} &     0    &    \textbf{0.008} &     0    &       0     &        0    \\
 P$_7$    &     0    &    0     &     0    &   0.0005 &     0.01 &    0.08  &     0.2  &   0.55   &     1    &    0.52  &     0.3  &    0.11  &     0.05 &       0.011 &        0    \\
 B$_8$    &     0    &    0     &     0    &   \textbf{0.008}  &     0    &    \textbf{0.051} &     0    &   \textbf{0.044}  &     0    &    1     &     0    &    \textbf{0.044} &     0    &       \textbf{0.05}  &        0    \\
 P$_8$    &     0    &    0     &     0    &   0      &     0    &    0.001 &     0.01 &   0.09   &     0.28 &    0.75  &     1    &    0.5   &     0.27 &       0.05  &        0.01 \\
 B$_9$    &     0    &    0     &     0    &   \textbf{0.002}  &     0    &    \textbf{0.016} &     0    &   \textbf{0.05}   &     0    &   \textbf{ 0.044} &     0    &    1     &     0    &       \textbf{0.044} &        0    \\
 P$_9$    &     0    &    0     &     0    &   0      &     0    &    0     &     0    &   0.001  &     0.05 &    0.1   &     0.28 &    0.75  &     1    &       0.4   &        0.1  \\
 B$_{10}$ &     0    &    0     &     0    &   0      &     0    &    0     &     0    &   0      &     0    &    \textbf{0.05 } &     0    &    \textbf{0.044} &     0    &       1     &        0    \\
 P$_{10}$ &     0    &    0     &     0    &   0      &     0    &    0     &     0    &   0      &     0    &    0     &     0.05 &    0.1   &     0.24 &       0.55  &        1    \\
\hline
\end{tabular}

  \captionof{table}{\textbf{Virtual matrix used in the experiment.} We highlight the newly introduced next-nearest barrier-barrier elements that allowed for parallel operation in \textbf{bold}. }
    \label{ED:matrix}

\vspace{2cm}
    
\centering\begin{tabular}{|l|c|c|c|c|}
\hline
\textbf{Gate} & \textbf{Fidelity} & \textbf{Magnetic noise} & \textbf{Charge noise} & \textbf{Other effects} \\
\hline
Sequential CNOT & $97.55 \pm 0.47\%$ & 1.03\% & 0.48\% & 0.94\%\\
Parallelized CNOT & $96.70\pm 0.57\%$ & 0.48\% & 0.45\% & 2.37\%\\
\hline
Sequential iSWAP & $97.43 \pm 0.43\%$ & 1.22\% & 0.56\% & 0.79\%\\
Parallelized iSWAP & $96.15\pm 0.57\%$ & 0.43\% & 0.56\% & 2.86\%\\
\hline
Sequential SWAP & $99.03 \pm 0.33\%$ & 0.17\% & 0.14\% & 0.66\%\\
Parallelized SWAP & $98.48 \pm 0.55\%$ & 0.05\% & 0.13\% & 1.34\%\\
\hline
\end{tabular}

  \captionof{table}{\textbf{Two-qubit RB error budget.}~Comparison of the two-qubit gate performance between sequential and parallel pulsing, including estimated error contributions from magnetic noise, charge noise and other effects. }
    \label{ED:comparison_table}

\newpage


\section*{Supplementary material}

# SUPPLEMENTARY MATERIALS
## Operating two exchange-only qubits in parallel

Mateusz T. Mądzik[*,†,1], Florian Luthi[†,1], Gian Giacomo Guerreschi[1], Fahd A. Mohiyaddin[1], Felix Borjans[1], Jason D. Chadwick[1], Matthew Curry[1], Joshua Ziegler[1], Sarah Atanasov[1], Peter L. Bavdaz[1], Elliot J. Connors[1], J. Corrigan[1], H. Ekmel Ercan[1], Robert Flory[1], Hubert C. George[1], Benjamin Harpt[1], Eric Henry[1], Mohammad M. Islam[1], Nader Khammassi[1], Daniel Keith[1], Lester F. Lampert[1], Todor M. Mladenov[1], Randy W. Morris[1], Aditi Nethwewala[1], Samuel Neyens[1], René Otten[1], Linda P. Osuna Ibarra[1], Bishnu Patra[1], Ravi Pillarisetty[1], Shavindra Premaratne[1], Mick Ramsey[1], Drew Risinger[1], John Rooney[1], Rostyslav Savytskyy[1], Thomas F. Watson[1], Otto K. Zietz[1], Anne Matsuura[1], Stefano Pellerano[1], Nathaniel C. Bishop[1], Jeanette Roberts[1], and James S. Clarke[1]

[1]Intel Corporation, Technology Research Group, Hillsboro, OR 97124, USA

April 1, 2025

## Contents

---

[*]mateusz.madzik@intel.com
[†]These authors contributed equally

# 1 Exchange-Only qubits

There are several ways to define qubits using electron spins in silicon quantum dot arrays. Here we provide more details related to the so-called "exchange-only qubit" which requires 3 electron spins each located in a different dot [1]. Since every electron has spin 1/2, we have a total of 8 states that are usually described in terms of the Z component of the individual dots as $\{|\uparrow\uparrow\uparrow\rangle, |\uparrow\uparrow\downarrow\rangle, |\uparrow\downarrow\uparrow\rangle, |\uparrow\downarrow\downarrow\rangle, |\downarrow\uparrow\uparrow\rangle, |\downarrow\uparrow\downarrow\rangle, |\downarrow\downarrow\uparrow\rangle, |\downarrow\downarrow\downarrow\rangle\}$. However, for the definition of the EO (exchange-only) qubit states, it is useful to identify the spin states using three different quantum numbers, namely the combined spin of the first two electron spins $S_{12}$, the total spin $S = S_{123}$, and its projection along the Z direction $M = Z_{123}$. Since these three observables commute, we can choose a basis of common eigenstates to describe the Hilbert space of the spin system: $\{|s_{12}, s, m\rangle\}_{s_{12}, s, m}$. Table 1 shows the possible combination of the three quantum numbers and identifies the qubit states. Intuitively, $s = 1/2$ identifies the qubit subspace, $s_{12} \in \{0, 1\}$ corresponds to the qubit state in the computational basis, $m$ acts as a gauge.

|  | $S_{12}$ | $S$ | $M$ | spin state |
|---|---|---|---|---|
| qubit $|0\rangle$ | 0 | 1/2 | 1/2 | $|S\rangle|\uparrow\rangle$ |
|  | 0 | 1/2 | -1/2 | $|S\rangle|\downarrow\rangle$ |
| qubit $|1\rangle$ | 1 | 1/2 | 1/2 | $\sqrt{\frac{2}{3}}|T_+\rangle|\downarrow\rangle - \sqrt{\frac{1}{3}}|T_0\rangle|\uparrow\rangle$ |
|  | 1 | 1/2 | -1/2 | $-\sqrt{\frac{2}{3}}|T_-\rangle|\uparrow\rangle + \sqrt{\frac{1}{3}}|T_0\rangle|\downarrow\rangle$ |
| leakage | 1 | 3/2 | 3/2 | $|T_+\rangle|\uparrow\rangle$ |
|  | 1 | 3/2 | 1/2 | $\sqrt{\frac{1}{3}}|T_+\rangle|\downarrow\rangle + \sqrt{\frac{2}{3}}|T_0\rangle|\uparrow\rangle$ |
|  | 1 | 3/2 | -1/2 | $\sqrt{\frac{1}{3}}|T_-\rangle|\uparrow\rangle + \sqrt{\frac{2}{3}}|T_0\rangle|\downarrow\rangle$ |
|  | 1 | 3/2 | -3/2 | $|T_-\rangle|\downarrow\rangle$ |

**Supplementary Data Table 1** | Definition of qubit and leakage states using the shorthands for singlet/triplet states $|S\rangle = (|\uparrow\downarrow\rangle - |\downarrow\uparrow\rangle)/\sqrt{2}$, $|T_0\rangle = (|\uparrow\downarrow\rangle + |\downarrow\uparrow\rangle)/\sqrt{2}$, $|T_+\rangle = |\uparrow\uparrow\rangle$, $|T_-\rangle = |\downarrow\downarrow\rangle$.

While it is tempting to think in terms of a "tensor product structure" like $|\psi\rangle = |s\rangle \otimes |s_{12}\rangle \otimes |m\rangle$, this picture is wrong since the Hilbert space has a "direct sum structure" with respect to the chosen quantum numbers. Similarly, it is tempting to think of the first two spins as determining the qubit state (after all it corresponds to $s_{12}$ for non-leaked states) and of the third spin as being associated only with the gauge, however this is again an incorrect picture since all three spins play an essential role in the qubit definition.

Note that all non-qubit states have $s = 3/2$ and they are usually referred to as leaked states (or just leakage for brevity). One should not confuse the definition of leakage in the context of EO qubits with other uses of that term, often related to the presence of an environment in addition to the system of interest. Finally, it is interesting to notice that $s_{12} = 1$ for all leaked states, so that their readout signature (obtained by measuring the first two spins in the singlet/triplet basis) is the same as that of the qubit $|1\rangle$ state.

Qubit operations are implemented by switching on/off the exchange interactions between pairs of spins. Intuitively, by lowering the potential barrier separating electrons in nearby wells, their spatial wavefunctions expand and overlap, increasing their spin-spin coupling. The interaction has the form $H_{jk}^{\text{exchange}} = J_{jk} S_j \cdot S_k$ and commutes with $S$ and $M$. This means that the exchange interaction between any pair of the three dots forming an EO qubit will never evolve a qubit state into a leaked state, but may instead drive a non-trivial evolution inside the qubit subspace. One can express the result of an exchange interaction of cumulative strength $J$ (i.e. the integration of $J_{ij}$ for the duration of the pulse) in terms of qubit operations via [1]:

- **[$j, k = 1, 2$]** Apart from global phases, this corresponds to a rotation around the Z axis of the qubit, namely $J_z(J) = R_z(-J) = \exp(iJZ/2)$

- **[$j, k = 2, 3$]** Apart from global phases, this corresponds to a rotation around the $n = -\sqrt{3}X/2 - Z/2$ axis of the qubit, namely $J_n(J) = R_n(-J) = \exp(-iJ(\sqrt{3}X + Z)/4)$

By concatenating a few $J_z$ and $J_n$ pulses, one can generate any 1Q gate. If one is interested in 1Q Clifford gates, at most 4 pulses are needed [2].

The situation is different for 2Q gates. Although they can be achieved via sequential exchange pulses between pairs of the 6 dots composing the 2 EO qubits, it is not guaranteed that the spin state remains in the qubit subspace during the

---

[1]On the axis of rotation, one has $S_1.S_2|_{\text{qubit}} = -I/4 + 1/2(-Z)$, $S_2.S_3|_{\text{qubit}} = -I/4 + 1/2(\sqrt{3}X/2 + Z/2)$, $S_1.S_3|_{\text{qubit}} = -I/4 + 1/2(-\sqrt{3}X/2 + Z/2)$

sequence. In fact, it typically does not because the leakage subspace is populated through exchange interactions involving dots from different qubits. When we speak of "exchange pulse sequences" to implement 2Q gates, such sequences must remove all intermediate leakage and, when executed in their entirety, map qubit states to qubit states in accordance with the desired operation.

Many sequences result in the same 2Q operation, and one may filter them based on the number of pulses or total duration (considered as proxies of the gate fidelity), or due to constraints like the dot connectivity. A recent work explores sequences for the CNOT, SWAP, iSWAP gates for multiple planar connectivities, adding the possibility of incorporating a desired permutation within the spins forming the same EO qubit [3].

## 2 Cross-Entropy Benchmarking

Cross-Entropy Benchmarking (XEB) was introduced to estimate the fidelity of multiqubit devices [4]. It relies on the ability to run instances of Random Quantum Circuits (RQC) of a particular ansatz designed to approximate Haar-random unitaries. One then checks how close the observed frequencies are to the expected probabilities by computing their cross-entropy (thus the name of this technique) or a linearized version of it. From the (linearized) cross-entropy one can infer the average fidelity of the RQCs and, by extension, of the device operations. The XEB played an important role in the validation of Google's quantum supremacy experiment [5] by confirming that the Sycamore device was within the error bounds believed to separate even the best classical approximations from the noiseless quantum probability distribution.

More formally, consider an ensemble $\{U\}$ of unitaries drawn according to a certain distribution. Ideally, this distribution would be the uniform distribution of unitaries according to the Haar measure. In practice, one proposes a certain ansatz and numerically verifies that it approximates Haar-random unitaries, for example, by comparing a few of its momenta. We discuss our ansatz further below in this section.

Given the unitary $U$, consider the state $|\psi_U\rangle$ obtained by applying $U$ to $|00\ldots0\rangle$. One can define the probability distribution expected when $|\psi_U\rangle$ is measured in the computational basis as:

$$p_U(x) = |\langle x|\psi_U\rangle|^2 = |\langle x|U|00\ldots0\rangle|^2 \quad (1)$$

In actual experimental implementations, decoherence and errors corrupt the state and affect the measurement process, so that the observed frequency of the possible outcomes, here denoted by $p_U^{\text{exp}}$, computed from $m$ experimental runs and the corresponding samples $S_{\text{exp}} = \{x_1^{\text{exp}}, x_2^{\text{exp}}, \ldots, x_m^{\text{exp}}\}$ does not necessarily agree with $p_U$.

A way to quantify how $p_U^{\text{exp}}$ differs from $p_U$ is the Kullback-Leibler divergence or, equivalently, apart from a term dependent on $p_U^{\text{exp}}$ alone, the cross entropy $H(p_U^{\text{exp}}, p_U) = -\sum_{j=1}^{D} p_U^{\text{exp}}(x_j)\log(p_U(x_j))$. Prior analysis strongly suggests that no classical algorithm, given $U$, can efficiently generate a distribution $P_U^{\text{alg}}$ closer to $p_U$ than the uniform distribution $p_{\text{uni}}(x) = 1/D$, which by its nature is uncorrelated with $p_U$. Of course, inefficient classical methods can sample from $p_U$ (for example by simulating the RQC), but their exponential computational cost limits their applicability to small systems, thus supporting the quantum supremacy claim.

Following Ref. [5], the cross-entropy benchmarking results are expressed via the linearized XEB fidelity. For $n$-qubit circuits, it is defined as

$$F_{\text{XEB}} = D\langle p_U(x_j)\rangle_j - 1 \quad (2)$$

with $D = 2^n$ and the angle brackets indicating the average over the observed bitstrings (here indicated with $x_j$). $F_{\text{XEB}}$ does not behave as a usual "fidelity" of quantum operations, since it may take values outside $[0, 1]$. However, since $F_{\text{XEB}} = 0$ when the observed $x_j$ are distributed uniformly and $F_{\text{XEB}} = 1$ when the $x_j$ are distributed according to $p_U$ itself, in practice $F_{\text{XEB}}$ is a good proxy of circuit fidelity when the effective noise model corresponds to qubit depolarization.

### 2.1 RQC ansatz

We draw inspiration from the RQC ansatz used in Google's supremacy paper [5] and maintain several of their features while adapting a few others to the EO systems.

- The RQC are formed by cycles of one- and two-qubit operations. Every cycle has random one-qubit operations on every qubit followed by a pattern of two-qubit gates.

- The one-qubit operations are chosen between the gates that are native to EO systems, namely $J_n$ and $J_z$ pulses. For each individual pulse, its rotation angle is chosen at random in $\{k\,\pi/4\}_{k=0,1,\ldots,7}$. As a comparison, in Ref. [5] the one-qubit operations are taken from $\{\sqrt{X}, \sqrt{Y}, \sqrt{W}\}$.

- The two-qubit gates are fixed and their pattern is one of 4 possible options, labeled A, B, C, and D. The sequence of the 2Q-gate patterns is fixed to ABCD CDAB and then repeated.

3

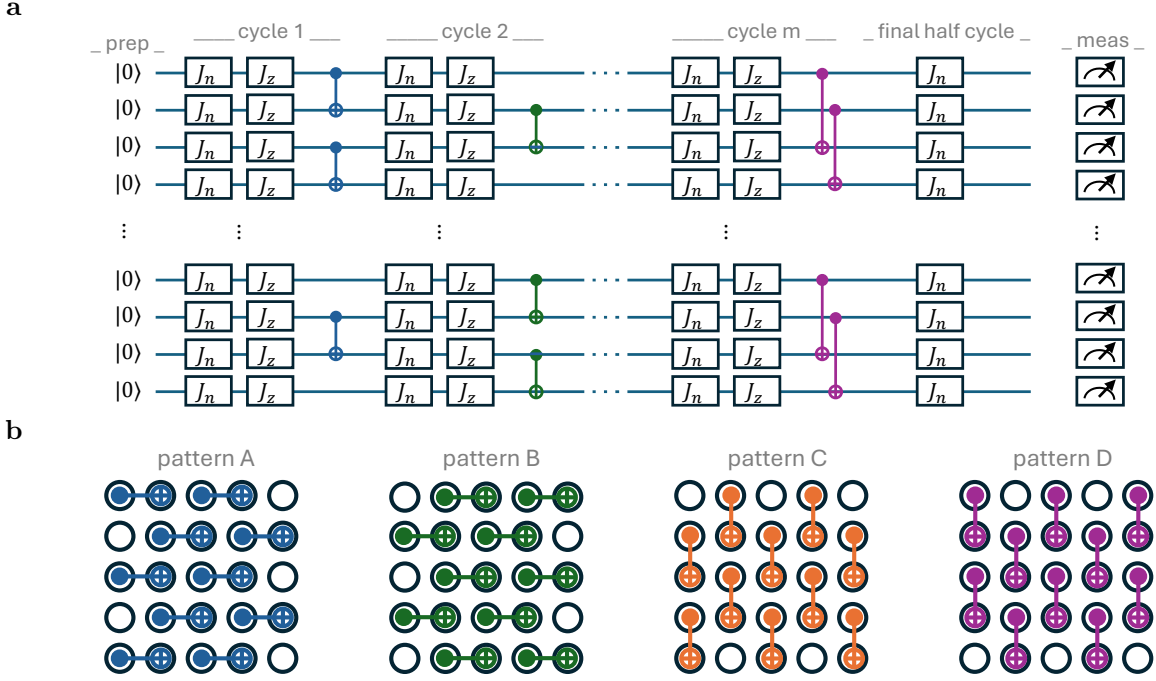

**Supplementary Figure 1 | a:** Random Quantum Circuit (RQC) based on $m$ cycles. Each cycle is formed by one-qubit $J_n$ and $J_z$ operations with random J values in $\{k\pi/4\}_{k=0,1,\ldots,7}$, followed by two-qubit CNOT gates. An additional half-cycle before measurements contains only $J_n$ pulses. **b:** Possible patterns of the CNOT gates for a 2D qubit array. The RQCs have cycles with patterns ABCD CDAB, possibly repeated as needed. For linear qubit arrays, we skip the patterns which have no two-qubit gate. Thus, in case of only two qubits, we are left solely with the pattern A, adopted in every cycle.

- Every two-qubit gate is a CNOT, different from the $\sqrt{\text{iSWAP}}$ gates in Ref. [5].

In case of small or unidimensional qubit arrays, we skip the 2Q-gate patterns that contain no CNOT gate. In particular, for any 1D array, the pattern sequence becomes AB AB. For a two-qubit RQC, each cycle is reduced to random $J_n$ and $J_z$ pulses followed by the same CNOT.

The caption of Supplementary Figure 1 describes the RQC ansatz in detail. We note that our experiments involve only 2 EO qubits. While this makes the generalization to larger RQCs (especially those related to 2D qubit arrays) unnecessary for the description of our experiment, we believe it is important to confirm that the two-qubit RQC can be seen as small-scale implementations of Haar-random multiqubit unitaries.

In Supplementary Figure 2 we show two visual ways to confirm the convergence of our RQC ansatz to Haar-random unitaries. In the left panel, we plot the histogram of all the outcome probabilities across 10 RQC instances ($4 \times 5$ qubit array), and compare it to the Porter-Thomas distribution. In the right panel, we changed the variable to $z = \log(Dp)$, with $D = 2^n$, to facilitate the comparison as in Ref. [5].

## 2.2 Peculiarity of EO decoherence

EO qubits are defined as collective states of 3 spins. Decoherence (usually associated with $T_2$ effects) and relaxation (related to $T_1$ effects) typically act at the level of the physical systems, here the electron spins, and their effect on the qubit state may differ from the common intuition.

In particular, decoherence during a long pulse sequence typically leads to the spins being fully depolarized, and this corresponds to every EO qubit being in the qubit $|0\rangle$ state with probability 25%, in the qubit $|1\rangle$ state with probability 25%, and in a leaked state with probability 50%. Since the readout signature of the leaked states is the same as the signature of $|1\rangle$, in practice the probability of assigning the result "1" increases to 75%. For this reason we introduce the probability distribution $p_{\text{ham}}$ which depends on the Hamming weight of the $n$-bit string:

$$p_{\text{ham}}(x) = \frac{3^{|x|}}{4^n} \quad (3)$$

where $|x|$ is the Hamming weight of $x$. In the XEB context, $n$ corresponds to the number of qubits in the RQC.

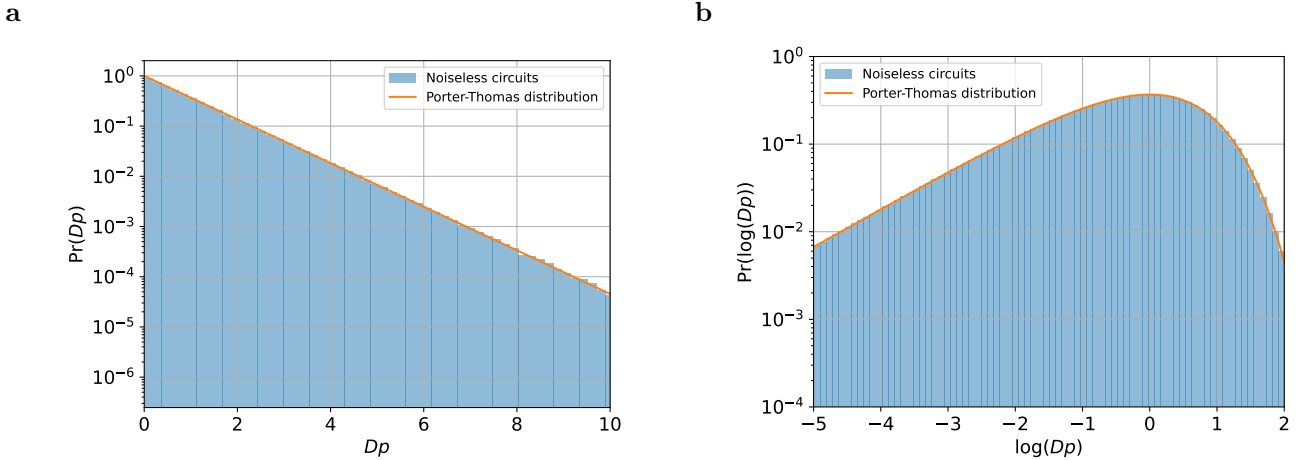

**Supplementary Figure 2** | Probability distribution of the outcomes, averaged over 10 RQCs on a $4 \times 5$ array of qubits and with 20 cycles. The circuits follow the ansatz described in the Supplementary Section 2.1. **Left:** Distribution of the outcome probabilities from noiseless simulations. $D = 2^{4\times 5}$ is the number of possible outcomes. In orange, the Porter-Thomas distribution $\Pr(p) = De^{-Dp}$. **Right:** Distribution of $\log(Dp)$. In orange $\Pr(z) = e^{z-e^z}$, obtained using the substitution $z = \log(Dp)$.

This observation does not change the noiseless probability $p_U$, nor the best estimate that is accessible, at scale, to a classical computer. However, it affects the experimental probability that, for a given $U$, is intermediate between $p_U$ and $p_{\text{ham}}$, instead of approaching $p_{\text{uni}}$ when noise dominates. For a single RQC, the cross entropy between $p_U$ and $p_{\text{ham}}$ depends on whether $|\psi_U\rangle$ has a large overlap with high-Hamming-weight states. At the same time, $p_{\text{ham}}$ is fixed and therefore independent of $U$. We have $\mathbb{E}_U[H(p_{\text{ham}}, p_U)] = -\sum_{j=1}^{N} p_{\text{ham}}(x_j)\mathbb{E}_U[\log(p_U(x_j))] = -H_0$ where, following [5], we define $H_0 \equiv \mathbb{E}_U[\log p_U(x_j)] = \log N + \gamma$. Here, $\gamma$ is Euler's constant. The cross-entropy difference $\Delta H(p) \equiv H_0 - H(p, p_U)$ is expected (when averaging over Haar-random $U$) to be unity when $p = p_U$ and zero when $p$ is an uncorrelated distribution like $p_{\text{ham}}$ or $p_{\text{uni}}$.

## 3 Significance of Crosstalk Calibration With Simultaneous Pulses

In this section, we highlight the importance of crosstalk calibration when simultaneously pulsing adjacent electrodes. Supplementary Figure 3a illustrates a situation where two virtual barrier electrodes vB8 and vB6 are pulsed, considering virtual compensation with the nearest and second nearest plunger electrodes. In the case where each virtual barrier electrode is calibrated independently of the other barrier electrodes being pulsed, simultaneous pulsing will hence result in an "uncalibrated" error due to crosstalk.

To quantify the impact of crosstalk, we first employ electrostatic simulations to calculate the relative impact of the adjacent electrodes on a tunnel barrier between two quantum dots, and thereby the exchange coupling, when compared to the nearest barrier electrode (Supplementary Figure 3b). Considering (i) realistic scenario of pulsing virtual barrier electrodes by 150 mV, and (ii) a simplified virtualization matrix in Supplementary Figure 3c, we then estimate the uncalibrated change in the effective voltage on the electrode and the resulting exchange pulse error using methods employed in Ref. [2], as adjacent virtual barrier electrodes are simultaneously pulsed (Supplementary Figure 3d).

We note from Figure 4d that two-qubit gates have several instances (7, 9 and 3 for CNOT, iSWAP and SWAP respectively) with simultaneous pulsing of the second nearest barriers, where errors for each exchange pulse will be $\approx 1.0$ %. The total two-qubit gate errors from crosstalk will hence accumulate and will even exceed the total two-qubit errors with sequential pulses shown in Figure 4c, if the gates are not calibrated accounting for simultaneous pulsing. We further emphasize that fault tolerant quantum circuits with quantum error correction will rely on heavy parallelism for the exchange pulses between nearby dots. The methods used for correcting crosstalk errors discussed in this paper will therefore be necessary given the stringent requirements of CNOT errors $< 0.1$ % for error correction.

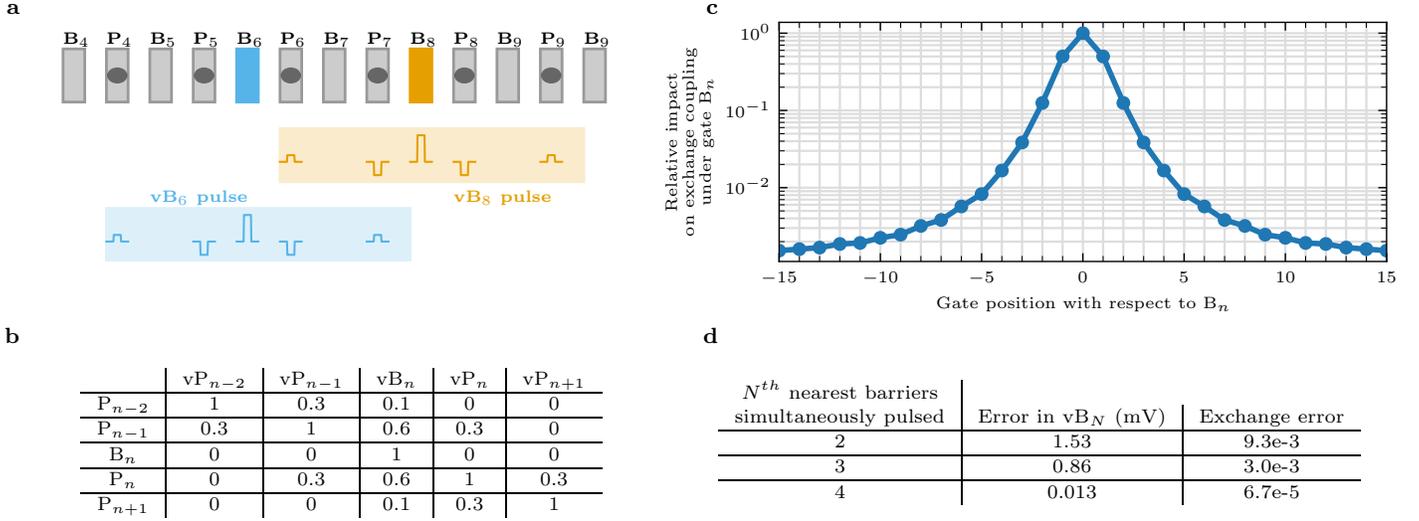

**Supplementary Figure 3** | (a) Illustration of physical electrode voltages when pulsing a virtual barrier. (b) Relative impact of the adjacent electrodes on the tunnel barrier under a barrier electrode, calculated with electrostatic simulations. (c) Virtualization matrix employed to estimate the resulting error in virtual barrier voltage, when adjacent barrier electrodes are simultaneously pulsed. (d) Error in virtual barrier voltage and resulting exchange pulse error, with simultaneous pulsing of nearby virtual barriers.



\end{document}